\documentclass[aps,prx,reprint,showpacs,
superscriptaddress,longbibliography,nofootinbib] {revtex4-2}
\usepackage{graphicx}
\usepackage{dcolumn}
\usepackage{bm}

\usepackage{afterpage}
\usepackage{float}
\usepackage{placeins}
\usepackage{dsfont}
\usepackage[utf8]{inputenc}
\usepackage{bm}
\usepackage{amssymb}
\usepackage{amsmath}
\usepackage{amsfonts}
\usepackage{graphicx}
\usepackage[usenames,dvipsnames]{xcolor} 
\usepackage{dcolumn}
\usepackage{bm}

\usepackage[bookmarks=true,colorlinks,citecolor=blue,urlcolor=blue]{hyperref}

\usepackage{bbm}
\usepackage{float}

\usepackage{dcolumn}   
\usepackage{bm}        
\usepackage{filecontents}
\usepackage{lineno}

\usepackage{mathtools}
\usepackage[usenames,dvipsnames]{xcolor} 

\usepackage[export]{adjustbox}
\usepackage{adjustbox}

\usepackage{braket}
\usepackage{tikz}
\usepackage{enumitem}

\newcommand\bs[1]{\ensuremath{\boldsymbol{#1}}}

\newcommand{\eq}[1]{Eq.\thinspace(\ref{#1})}
\newcommand{\eqs}[2]{Eqs.\thinspace(\ref{#1},\ref{#2})}
\newcommand{\eqss}[2]{Eqs.\thinspace(\ref{#1}-\ref{#2})}
\newcommand{\fig}[1]{Fig.\thinspace{}\ref{#1}}
\newcommand{\fc}[1]{({#1})}
\newcommand{\figc}[2]{Fig.\thinspace{}\ref{#1}\thinspace{}\fc{#2}}

\newcommand{\dwpair}{
\tikz[baseline=-0.62ex]{
\draw [line width=0.35mm,black] (0,0) -- (0.4,0); 
\node[mark size=2pt,color=black] at (0,0) {\pgfuseplotmark{*}};
\node[mark size=2pt,color=black] at (0.4,0) {\pgfuseplotmark{*}};
\draw[thick] (0,0) circle (2pt);
\draw[thick] (0.4,0) circle (2pt); }
}

\newcommand{\dwalpair}{
\tikz[baseline=-0.62ex]{
\node[mark size=2pt,color=black] at (0,0) {\pgfuseplotmark{*}};
\draw [black,thick,domain=35:145] plot ({0.24*cos(\x)+0.2}, {0.24*sin(\x)-0.08});
\draw[thick] (0,0) circle (2pt);
\draw[thick] (0.4,0) circle (2pt); }
}

\newcommand{\alsingle}{
\tikz[baseline=-0.62ex]{
\draw[thick] (0,0) circle (2pt); }
}

\newcommand{\plaqv}{
\tikz[baseline=-0.62ex]{
\draw [gray] (0,-0.19) -- (0.3,-0.19);
\draw [gray] (0,0.19) -- (0.3,0.19);
\filldraw [color=black, fill=black!65] (0,0) ellipse (0.06 and 0.23); 
\filldraw [color=black, fill=black!65] (0.3,0) ellipse (0.06 and 0.23); }
}

\newcommand{\plaqh}{
\tikz[baseline=-0.62ex]{
\draw [gray] (-0.19,-0.19) -- (-0.19,0.19);
\draw [gray] (0.2,-0.19) -- (0.2,0.19);
\filldraw [color=black, fill=black!65] (0,-0.15) ellipse (0.23 and 0.058); 
\filldraw [color=black, fill=black!65] (0,0.15) ellipse (0.23 and 0.058); }
}

\begin{document}

\title{Emergent tracer dynamics in constrained quantum systems}
\author{Johannes Feldmeier}
\affiliation{Department of Physics, Technical University of Munich, 85748 Garching, Germany}
\affiliation{Munich Center for Quantum Science and Technology (MCQST), Schellingstr. 4, D-80799 M{\"u}nchen, Germany}
\author{William Witczak-Krempa}
\affiliation{Université de Montréal, C.P. 6128, Succursale Centre-ville, Montréal, QC, Canada, HC3 3J7}
\affiliation{Institut Courtois, Universit\'e de Montr\'eal, Montr\'eal (Qu\'ebec), H2V 0B3, Canada}
\affiliation{Centre de Recherches Mathématiques, Université de Montréal, Montréal, QC, Canada, HC3 3J7}
\author{Michael Knap}
\affiliation{Department of Physics, Technical University of Munich, 85748 Garching, Germany}
\affiliation{Munich Center for Quantum Science and Technology (MCQST), Schellingstr. 4, D-80799 M{\"u}nchen, Germany}
\date{\today}

\begin{abstract}
We show how the tracer motion of tagged, distinguishable particles can effectively describe transport in various homogeneous quantum many-body systems with constraints. We consider systems of spinful particles on a one-dimensional lattice subjected to constrained spin interactions, such that some or even all multipole moments of the effective spin pattern formed by the particles are conserved. On the one hand, when all moments---and thus the entire spin pattern---are conserved, dynamical spin correlations reduce to tracer motion identically, generically yielding a subdiffusive dynamical exponent $z=4$. This provides a common framework to understand the dynamics of several constrained lattice models, including models with XNOR or $tJ_z$ -- constraints. We consider random unitary circuit dynamics with such a conserved spin pattern and use the tracer picture to obtain exact expressions for their late-time dynamical correlations. Our results can also be extended to integrable quantum many-body systems that feature a conserved spin pattern but whose dynamics is insensitive to the pattern, which includes for example the folded XXZ spin chain.
On the other hand, when only a finite number of moments of the pattern are conserved, the dynamics is described by a convolution of the internal hydrodynamics of the spin pattern with a tracer distribution function. As a consequence, we find that the tracer universality is robust in generic systems if at least the quadrupole moment of the pattern remains conserved.  In cases where only total magnetization and dipole moment of the pattern are constant, we uncover an intriguing coexistence of two processes with equal dynamical exponent but different scaling functions, which we relate to phase coexistence at a first order transition.
\end{abstract}

\maketitle

{
\hypersetup{linkcolor=black}
\tableofcontents
}
    
\section{Introduction} \label{sec:introduction}
Recent years have seen rapid progress in quantum simulation technology, with increasing capacity to directly probe the out-of-equilbrium properties of many-body quantum systems. These advances have led to immense theoretical and experimental interest in the thermalization of closed, interacting quantum many-body systems towards translationally invariant equilibrium states~\cite{Deutsch91, Srednicki94, Rigol2008, alessio2016_chaos, 
kaufman2016_thermalization, Brydges2019_renyi}. A breakthrough has been the key realization that the late time dynamics in such systems can be understood via an emergent, effectively classical hydrodynamic description. This includes the diffusive transport of local densities in systems with global conserved charges~\cite{chaikin_lubensky_1995,Mukerjee06,Lux14,Bohrdt16}, as well as 
the dynamics of entanglement~\cite{nahum2017_entanglement,jonay2018_coarse,knap2018_scrambling,
rakovszky2019_renyi,rakovszky2019_entanglement} and quantum information~\cite{nahum2018_operator,Keyserlingk2018,khemani2018_operator,
Rakovszky18,nahum2018_griffith,Parker19}. 

The remarkable emergence of classical hydrodynamics from a closed quantum time evolution is currently also being explored in many-body systems featuring more exotic conservation laws---or \textit{constraints}---such as 
gauge theories and fractonic quantum matter~\cite{nandkishore2019_fractons,pretko2020_fracton,
chamon2005_glass,haah2011_code,yoshida2013_fractal,vijay2015_topo,
Vijay16,pretko2018_elasticity,pretko2017_subdim,
pretko2018_gaugprinciple,pretko_witten,williamson2019_fractonic}. Crucially, constraints generally have a qualitative impact on the thermalization process of many-body systems towards equilibrium. While in some instances the presence of fractonic constraints can be as severe as precluding thermalization altogether~\cite{Sala19,khemani20192d,Rakovszky20,scherg2021_kinetic}, in many others they lead to novel subdiffusive universality classes of (emergent) hydrodynamic relaxation of the non-equilibrium evolution at late times~\cite{gromov2020_fractonhydro,feldmeier2020anomalous,
morningstar2020_kinetic,zhang_2020,Iaconis19,
feldmeier2021_fractondimer,moudgalya2021_spectral,Guardado20,
Singh2021_subdiff,iaconis2021_multipole,glorioso2021_breakdown,
grosvenor2021_hydro,osborne2022_fracton,feldmeier2021_critical,burchards2022_coupled}.

\begin{figure}[t]
\centering
\includegraphics[trim={0cm 0cm 0cm 0cm},clip,width=0.99\linewidth]{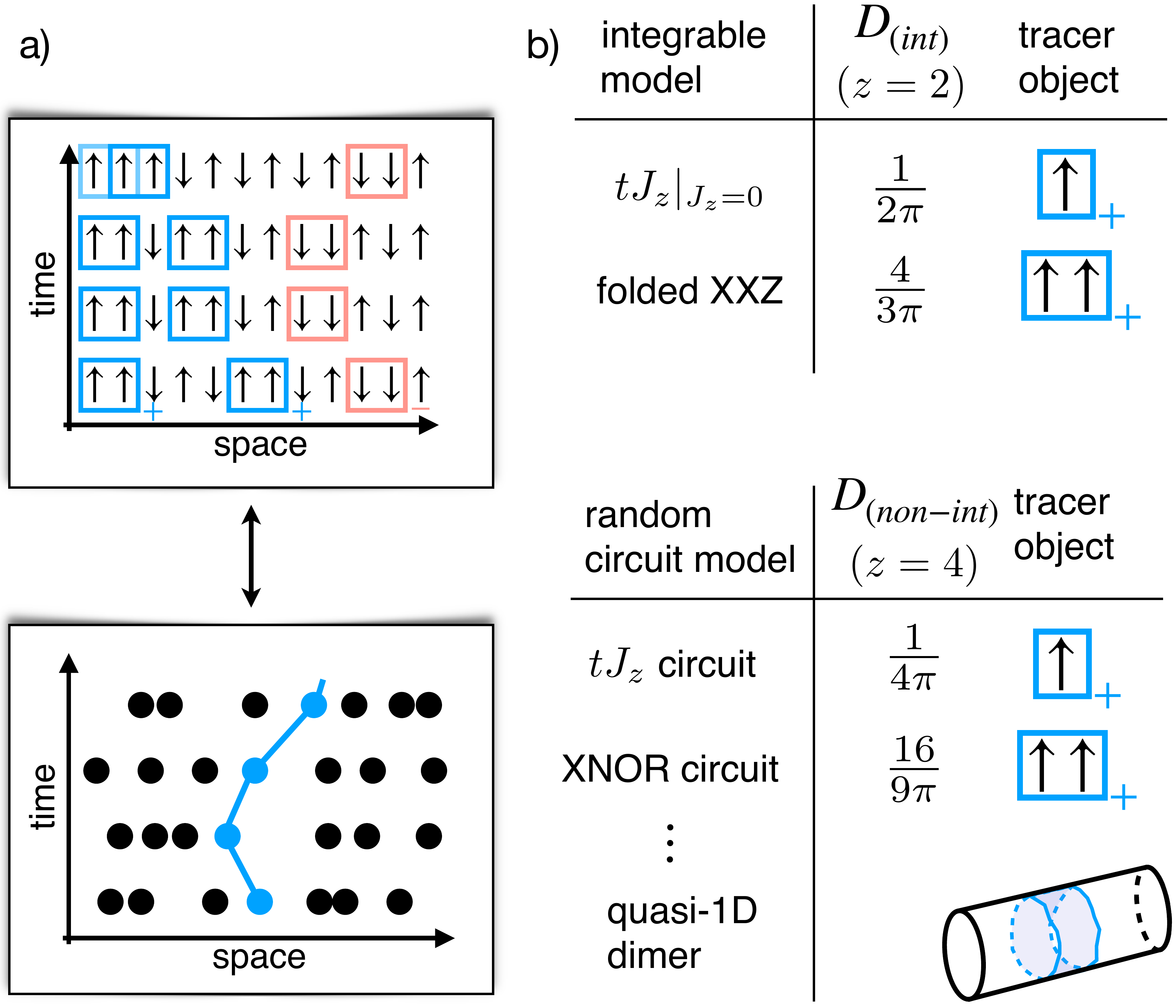}
\caption{\textbf{Tracer diffusion in constrained quantum systems.} \textbf{a)} We consider one-dimensional systems with a pattern of effective excitations (blue and red squares) which is conserved during the time evolution. Infinite temperature dynamical correlation functions in such an ensemble map directly on the tracer probability distribution of hard core random walkers. The fundamental objects of tracer diffusion are the effective excitations of the conserved pattern.  \textbf{b)} Solving the tracer problem provides us with quantitatively accurate descriptions of transport in a number of generic random unitary circuit models as well as integrable quantum systems. The effective conserved  patterns can assume a complex structure as in the quasi one-dimensional dimer model studied in Ref.~\cite{feldmeier2021_fractondimer}.}
\label{fig:1}
\end{figure}

In this work, we study the emergence of another classical process in the dynamics of interacting quantum many-body systems: the tracer motion of tagged particles. While at first sight the notion of a tagged particle appears to be at odds with the indistinguishability of quantum particles in many-body systems, here we show how the effects of kinetic constraints can nonetheless lead to the emergence of such tracer motion. For this purpose we focus on the dynamics of one-dimensional systems with a conserved pattern of effective spins or charges throughout much of this work, see \fig{fig:1} for an illustration. This setup is similar to certain nearest-neighbor simple exclusion processes in classical two-component systems, where tracer motion describes the local component imbalance~\cite{spohn2012_large}. Similar constraints have recently also been discussed in the context of fractonic quantum systems in terms of ``Statistically Localized Integrals of Motion'' (SLIOMs)~\cite{Rakovszky20}, which can be interpreted as an effective conserved pattern. 

More generally, we investigate the dynamics of local spin correlations in one-dimensional systems featuring a conserved number of spinful particles. The setup is similar to the $tJ$ -- model, which consists of spinful fermions with the condition of no double occupancies. In our case, the usual Heisenberg spin exchange is substituted by constrained spin interactions: We require that some or even all multipole moments of the spin pattern formed by the particles are conserved. For much of this work we focus on random unitary circuits that satisfy these constraints. We will therefore call the systems studied in this work `$tJ$ -- like'. We find that the anomalously slow tracer diffusion of hard core particles in one dimension plays a vital role in describing their dynamical spin correlations. 

The mapping between spin correlations and tracer dynamics becomes exact for systems with an exactly conserved spin pattern, where the tracer motion gives rise to a subdiffusive dynamical exponent $z=4$. Such systems are similar in structure to the $tJ_z$ -- model, where spin interactions diagonal in the $z$-basis preserve the spin pattern. We thus call such systems `$tJ_z$ -- like'.
This framework yields a unifying picture to understand the dynamics of constrained lattice models studied in recent works that can be mapped---either directly or effectively---to a $tJ_z$ -- like structure~\cite{Tomasi19,yang2020_strict,Rakovszky20,feldmeier2021_fractondimer,
Singh2021_subdiff}. 
We use this picture to derive the full long-time profile of the dynamical spin correlations in a random unitary $tJ_z$ -- circuit model and a random XNOR circuit~\cite{Singh2021_subdiff}.

Although our main focus is on the dynamics of generic systems, we demonstrate that the tracer picture is applicable also to certain integrable quantum systems. These feature an effective conserved spin pattern but their dynamics \emph{per se} is insensitive to this pattern. As examples we consider the integrable $J_z\rightarrow 0$ limit of the $tJ_z$ -- model and the folded XXZ chain~\cite{Zadnik2021_fxxz1,Zadnik2021_fxxz2,pozsgay2021_fxxz,
bidzhiev2022_fxxz}. Through the tracer picture we are able to reproduce their spin diffusion constants at infinite temperature and predict the full profile of their spin correlations at late time, in agreement with our numerical simulations.  

We then consider models in which only a finite number of moments of the spin pattern are conserved. The resulting spin correlations are given by a convolution of the tracer motion and the internal dynamics of the pattern. As a consequence, we find that the tracer-motion universality is robust to breaking the pattern conservation if all moments up to at least the quadrupole moment of the pattern are conserved. In addition, for dipole-conserving spin interactions we uncover a competition between two hydrodynamic processes that both have dynamical exponent $z=4$ but that exhibit different scaling functions. The long-time profile of the spin correlations is then described by a non-universal mixture of these two scaling functions. We argue that this intriguing situation is reminiscent to phase coexistence at a first order transition between a Gaussian and a non-Gaussian hydrodynamic phase.

The remainder of this paper is structured as follows: In Sec.~\ref{sec:models} we introduce the $tJ$ -- like models studied in this work and derive a general expression for their spin correlations at late times. We apply these results to specific random unitary circuit examples in Sec.~\ref{sec:generic}, treating in detail the random XNOR model~\cite{Singh2021_subdiff}. We consider two integrable models in Sec.~\ref{sec:integrable} and discuss cases where only a finite number of multipole moments of the pattern are conserved in Sec.~\ref{sec:multipole}.

\section{Models and spin correlations} \label{sec:models}
We introduce a novel class of $tJ$ -- like many-body systems of spinful particles in one dimension with constrained spin interaction terms. The constraints are such that either the entire spin pattern or a finite number of multipole moments of the pattern are conserved. We derive a general expression for the infinite temperature dynamical spin correlations at late times in such systems.
\subsection{Constrained $tJ$ -- like systems}
We are interested in conservation laws inherent to models of the form
\begin{equation} \label{eq:1.1}
\begin{split}
&\hat{H}^{}_m = -t \sum_{x=-L/2}^{L/2-1} \sum_{\sigma} (\hat{\tilde{c}}^\dagger_{x+1,\sigma}\hat{\tilde{c}}_{x,\sigma} + h.c.) + \hat{H}^{}_{S,m} \\
& \qquad \qquad [\hat{H}^{}_{S,m},\sum_x x^n \hat{S}^z_x] = 0 \quad \forall \; n\leq m.
\end{split}
\end{equation}
\eq{eq:1.1} describes a constant number of spinful fermions with nearest-neighbor hopping on a one-dimenional lattice, with the usual $tJ$ -- constraint of no double occupancies (indicated by the tilde over the fermion operators; the fermionic nature of the particles is not essential here). A spin pattern is then formed by the fermions in a squeezed space where all empty sites are removed. The spin interaction $\hat{H}_{S,m}$ between the fermions is generalized to not only conserve the total magnetization but potentially higher moments of this spin pattern (all up to the $m$th moment) as well. Examples for $\hat{H}_{S,m}$ include
\begin{equation} \label{eq:1.2}
\begin{split}
&\hat{H}^{}_{S,0} = J \sum_x \hat{\bs{S}}_{x} \cdot \hat{\bs{S}}_{x+1} + ... \\
&\hat{H}^{}_{S,1} = J \sum_x (\hat{S}^+_{x}\hat{S}^-_{x+1}\hat{S}^-_{x+2}\hat{S}^+_{x+3} + h.c.) + ... \\
&... \\
&\hat{H}^{}_{S,\infty } = J \sum_x \hat{S}^z_{x}\hat{S}^z_{x+1}+ ...,
\end{split}
\end{equation}
where `...' refers to diagonal terms in the $z-$basis or to longer-range off-diagonal terms that fulfill the conservation law of \eq{eq:1.1}. We note that $\hat{H}^{}_0 = \hat{H}^{}_{tJ}$ is a conventional $tJ$ -- model while $\hat{H}^{}_\infty = \hat{H}^{}_{tJ_z}$ is a $tJ_z$ -- model in which the entire spin pattern is a constant of motion. Lattice spin models such as \eq{eq:1.2} provide a novel way of interpolating between these two limiting cases; one can construct such models recursively~\cite{feldmeier2020anomalous}.

The Hamiltonians of \eqs{eq:1.1}{eq:1.2} serve as our starting motivation and we can qualitatively determine their universal late-time dynamics at high energies by considering \textit{generic} many-body systems with the same Hilbert space structure and conserved quantities. 
To introduce a model-independent notation we expand any state $\ket{\psi}$ with a fixed number $N_f$ of particles as $\ket{\psi} = \sum_{\bs{x},\bs{\sigma}} \psi(\bs{x},\bs{\sigma})\ket{\bs{x},\bs{\sigma}}$ in terms of the basis states
\begin{equation} \label{eq:1.3}
\ket{\bs{x},\bs{\sigma}}, \quad x_1 < ... < x^{}_{N_f}, \;\, \sigma_i \in \{\pm 1\}.
\end{equation}
Here, $\bs{x}$ labels the positions of the particles on the chain from left to right and $\bs{\sigma}$ their respective spins in the $z-$basis. The time evolution represented by the unitary $\hat{U}_m(t)$ should then fulfill
\begin{equation} \label{eq:1.4}
[\hat{U}_m(t),\sum_{j=1}^{N_f} j^n \hat{\sigma}_j] = 0 \quad \forall n \leq m,
\end{equation}
and can be Hamiltonian, such as in \eq{eq:1.1}, or generic, such as in random unitary quantum circuits or classical stochastic lattice gases; either case is expected to exhibit the same universal dynamical behavior. We emphasize that the conservation of moments in \eq{eq:1.4} applies to the squeezed-space variables of the pattern, which are related to the original spins non-locally. In particular, the moments $\sum_x x^n \hat{S}^z_x$ in the original spin space are \textit{not} conserved due to the hopping part of \eq{eq:1.1}. We will make use of the non-local property \eq{eq:1.4} throughout our work and show that various constrained models studied recently are part of this effective description.

\subsection{Generic structure of dynamical spin correlations}
We derive a general expression for the dynamical spin correlations in a system described by \eqs{eq:1.3}{eq:1.4} under the assumption of chaotic, thermalizing dynamics at infinite temperature. Integrable dynamics will be considered in Sec.~\ref{sec:integrable}.  We will assume open boundary conditions in a system of length $L$ containing a fixed number of $N_f$ particles, i.e. density $\rho=N_f/L$. The spin operator $\hat{S}^z_r$ at site $r$ can then be expressed in terms of the pattern spin operators $\hat{\sigma}_j$ via
\begin{equation} \label{eq:2.1}
\hat{S}^z_r = \sum_{j=1}^{N_f} \, \delta_{\hat{x}_j,r}^{} \, \hat{\sigma}_j.
\end{equation}
The time evolution $\hat{U}_m(t)$ applied to a basis state $\ket{\bs{x},\bs{\sigma}}$ is given by
\begin{equation} \label{eq:2.2}
\hat{U}_m(t)\ket{\bs{x},\bs{\sigma}}=\sum_{\bs{x}^\prime,\bs{\sigma}^\prime}a(\bs{x}^\prime\bs{\sigma}^\prime|\bs{x}\bs{\sigma};t)\ket{\bs{x}^\prime,\bs{\sigma}^\prime},
\end{equation} 
with the matrix elements $a(\bs{x}^\prime\bs{\sigma}^\prime|\bs{x}\bs{\sigma};t)$ normalized to $\sum_{\bs{x}^\prime,\bs{\sigma}^\prime} |a(\bs{x}^\prime\bs{\sigma}^\prime|\bs{x}\bs{\sigma};t)|^2=1$. Using \eqs{eq:2.1}{eq:2.2}, the dynamical spin correlations read
\begin{equation} \label{eq:2.3}
\begin{split}
&C(r,t) := \braket{\hat{S}^z_{r}(t)\hat{S}^z_{0}(0)} = \\
&\quad = \frac{1}{\mathcal{N}} \sum_{\substack{\bs{x},\bs{\sigma},i \\ \bs{x}^\prime,\bs{\sigma}^\prime,j}} \sigma^\prime_j \sigma_i \, \delta_{x^\prime_j,r}\delta_{x_i,0} \, |a(\bs{x}^\prime\bs{\sigma}^\prime|\bs{x}\bs{\sigma};t)|^2,
\end{split}
\end{equation}
where the normalization $\mathcal{N} = \mathcal{N}_n \mathcal{N}_s$ is given by the number of different particle position $\mathcal{N}_n = \binom{L}{N_f}$ on the lattice and the number of different spin patterns $\mathcal{N}_s = 2^{N_f}$. The expectation value $\braket{\cdot}$ is taken with respect to an `infinite temperature' ensemble over all basis states.
Under $\hat{U}_m(t)$, both the number of particles as well as the total magnetization of the spin pattern are conserved and we expect both of their local densities to contribute a hydrodynamic mode at long length scales and late times. In general, the precise transport coefficients of the particle mode and the spin pattern mode are determined by a mode-coupled Ansatz and the details of the microscopic time evolution. Nonetheless, we expect that \textit{qualitatively}, we can describe the hydrodynamic behavior of the two modes independently at late times in thermalizing systems. We therefore make the approximation to set
\begin{equation} \label{eq:2.4}
|a(\bs{x}^\prime\bs{\sigma}^\prime|\bs{x}\bs{\sigma};t)|^2 \simeq p_{n}(\bs{x}^\prime|\bs{x};t) \, p_{s}(\bs{\sigma}^\prime|\bs{\sigma};t)
\end{equation}
in \eq{eq:2.3}, where we introduced the particle and spin path distributions $p_{n}(\bs{x}^\prime|\bs{x};t)$ and $p_{s}(\bs{\sigma}^\prime|\bs{\sigma};t)$; they fulfill $\sum_{\bs{x}^\prime}p_{n}(\bs{x}^\prime|\bs{x};t) = 1 = \sum_{\bs{\sigma}^\prime}p_{s}(\bs{\sigma}^\prime|\bs{\sigma};t)$. The spin correlations of \eq{eq:2.3} are thus determined by the following two expressions which describe spin pattern dynamics and particle dynamics, respectively:
\begin{equation} \label{eq:2.5}
\begin{split}
F(j-i,t) &:= \frac{1}{\mathcal{N}_s}\sum_{\bs{\sigma},\bs{\sigma}^\prime} \sigma^\prime_j \sigma_i \, p_{s}(\bs{\sigma}^\prime|\bs{\sigma};t) \\
K(j-i,r;t) &:= \frac{1}{\mathcal{N}_n}\sum_{\bs{x},\bs{x}^\prime} \delta_{x^\prime_j,r}\delta_{x_i,0} \, p_{n}(\bs{x}^\prime|\bs{x};t) = \\
&= P(j,r|i,0;t) \, P(i,0).
\end{split}
\end{equation}
In the last step we introduced the probability $P(i,0)$ to find the $i$th particle (counted from the left) at site $x=0$, as well as the probability $P(j,r|i,0;t)$ to find the $j$th particle at site $x=r$ at time $t$ given that particle $i$ was located at site $x=0$ at time $0$. We can rewrite the latter probability as
\begin{equation} \label{eq:2.6}
P(j,r|i,0;t) = \sum_{\ell} P(j,r|i,\ell;0)\, P(i,\ell|i,0;t).
\end{equation}
We notice that $P(j,r|i,\ell;0)$ in \eq{eq:2.6} is simply the probability to find particle $j$ at $r$ given that particle $i$ is at $\ell$ at the same time. It has the exact expression ($\theta(\cdot)$ is the Heaviside theta function)
\begin{equation} \label{eq:2.7}
\begin{split}
&P(j,r|i,\ell;0) = \,\delta_{i-j,0} \, \delta_{r-\ell,0} \, + \\
& \quad + \theta(j-i-1)\theta(r-\ell-1)\frac{ \binom{r-\ell-1}{j-i-1} \binom{L-r+\ell-1}{N-j+i-1} }{ \binom{L}{N} } + \\
& \quad + \theta(i-j-1)\theta(\ell-r-1)\frac{ \binom{\ell-r-1}{i-j-1} \binom{L-\ell+r-1}{N-i+j-1} }{ \binom{L}{N} } \\
&\stackrel{|i-j|\gg 1}{\approx} \frac{1}{\sqrt{\pi \, \bigl(\frac{1}{\rho^2}-\frac{1}{\rho}\bigr)\, |j-i|}}\exp\Biggl\{ -\frac{\bigl[r-\ell-\frac{j-i}{\rho}\bigr]^2}{\bigl(\frac{1}{\rho^2}-\frac{1}{\rho}\bigr)\,|j-i|} \Biggr\},
\end{split}
\end{equation}
where in the last line we made an approximation for large $|i-j|\gg 1$, leading to a Gaussian centered around $r-\ell-\frac{j-i}{\rho}=0$ with width proportional to $\sqrt{|i-j|}$.\footnote{The last line of \eq{eq:2.7} actually follows in a grand canonical setting with average density $\rho$ of particles. Nonetheless, the location of the center and the scaling of the width remain valid for fixed particle number.} For the second expression on the right hand  side of \eq{eq:2.6} we define
\begin{equation} \label{eq:2.8}
P(i,\ell|i,0;t) =: G_{tr}(\ell,t),
\end{equation}
since $P(i,\ell|i,0;t)$ traces the motion of particle $i$, which we assume to be in the bulk of the spin pattern. $G_{tr}(\ell,t)$ thus corresponds to the time dependent tracer probability distribution of a bulk particle. Using \eqs{eq:2.7}{eq:2.8} in \eqs{eq:2.5}{eq:2.6} we obtain
\begin{equation}  \label{eq:2.9}
\begin{split}
K(j-i,r;t) &= P(i,0) \,\int d\ell \, P(j,r|i,\ell;0) \, G_{tr}(\ell,t) \simeq \\
&\simeq P(i,0) \, G_{tr}\Bigl(r-\frac{j-i}{\rho},t\Bigr).
\end{split}
\end{equation}
The last line follows since $G_{tr}(\ell,t)$ is in general a probability distribution whose width increases in time while the width of $P(j,r|i,\ell;0)$ is a constant of order $\sqrt{|j-i|}$. Therefore, at late times $G_{tr}(\ell,t)$ is much broader and we can substitute the approximation $P(j,r|i,\ell;0) \simeq \delta(r-\ell-\frac{j-i}{\rho})$ into \eq{eq:2.9}. Finally, inserting \eq{eq:2.9} and \eq{eq:2.5} into \eq{eq:2.3} we find
\begin{equation} \label{eq:2.10}
C(r,t) = \int d j \, F(j,t) \, G_{tr}(r-j/\rho,t) = [\tilde{F} \, \star \, G_{tr} ](r,t),
\end{equation}
with $\tilde{F}(j,t) = \rho \, F(\rho j,t)$. The dynamical spin correlations $C(r,t)$ at late times are thus given quite generally as a convolution between the internal dynamics of the spin pattern and the tracer distribution of a distinguishable particle on the lattice. Since $G_{tr}(\ell,t)$ follows the trajectory of the $i$th particle (with some $i$ in the bulk) counted from the left, the relevant tracer problem is one of hard core interacting particles that can never swap relative positions. We note that in reciprocal space \eq{eq:2.10} represents two independent decay processes of long wavelength $k$-modes.

\section{Random unitary circuits with conserved pattern} \label{sec:generic}
We use the result of \eq{eq:2.10} to study a number of $tJ_z$ -- like models with a spin pattern that is a constant of motion. We remark that the results of this section should apply very generally to models featuring recently introduced ``Statistically Localized Integrals of Motion'' (SLIOMs)~\cite{Rakovszky20}, which can be interpreted as a conserved pattern.

If the entire pattern is constant, the spin dynamics becomes trivial, $F(j,t)=\delta(j)$ for all times in \eq{eq:2.10} and thus 
\begin{equation} \label{eq:3.0.0}
C(r,t) \simeq G_{tr}(r,t)
\end{equation}
maps directly to a tracer problem.
Due to the trivial pattern dynamics, our initial approximation \eq{eq:2.4} simply becomes $|a(\bs{x}^\prime\bs{\sigma}|\bs{x}\bs{\sigma};t)|^2 \simeq p_n(\bs{x}^\prime|\bs{x};t)$, i.e., the matrix elements of the time evolution can be considered approximately independent of the underlying spin pattern when inserted into \eq{eq:2.3}. In fact, with a conserved pattern the correlations of \eq{eq:2.3} can be recast as
\begin{equation} \label{eq:3.0.1}
\begin{split}
C(r,t) = G_{\mathrm{tr}}(r,t) + R(r,t),
\end{split}
\end{equation}
where $R(r,t)$ is the difference between the \textit{exact} correlations and the tracer distribution. It reads explicitly
\begin{equation} \label{eq:3.0.2}
\begin{split}
&R(r,t) = \frac{1}{\mathcal{N}} \sum_{\substack{\bs{x},\bs{x}^\prime, \bs{\sigma} \\ i\neq j}} \sigma_j\sigma_i \, \delta^{}_{x^\prime_j,r}\, \delta^{}_{x_i,0} |a(\bs{x}^\prime\bs{\sigma}|\bs{x}\bs{\sigma};t)|^2,
\end{split}
\end{equation}
and captures contributions to $C(r,t)$ due to spins $j\neq i$ moving to site $r$ at time $t$, given that spin $i$ started at site $x=0$ initially.
Due to the summation over the spin values $\sigma_j$, $R(r,t)$ acquires both a positive and negative contribution from $\sigma_j$ and $-\sigma_j$, respectively.
We then expect generically that contributions to $R(r,t)$ from spins $j$ with $|j-i| \gg 1$ vanish approximately due to cancellation of positive and negative contributions, justifying \eq{eq:3.0.0}. In this section, we will consider generic systems where $R(r,t)=0$ exactly upon averaging over the random time evolution, and in Sec.~\ref{sec:integrable} integrable quantum systems in which $R(r,t)=0$ exactly since the time evolution is indeed independent of the underlying pattern, hence in both cases $|a(\bs{x}^\prime\bs{\sigma}|\bs{x}\bs{\sigma};t)|^2 = p_n(\bs{x}^\prime|\bs{x};t)$. In either case, since $C(r,t)=G_{\mathrm{tr}}(r,t)$ exactly, we will be able to use existing results from the theory of tracer dynamics to obtain full long-time spin correlation profiles.

Here, we first consider systems subject to a random time evolution, such as a classical stochastic lattice gas or random unitary quantum circuits. Averaging over the random evolution (denoted by $\overline{\cdot\cdot\cdot}$) we are interested in the associated averaged correlations $\overline{C(r,t)}$. For certain models that we consider, $\overline{R(r,t)}=0$ (see below) and thus the mapping to tracer dynamics is exact upon averaging over the random evolution, $\overline{C(r,t)}=G_{\mathrm{tr}}(r,t)$. 
The resulting tracer problem we have to solve is one of particles hopping randomly  on a one-dimensional lattice subject to a hard core exclusion principle. The hard core property is a direct consequence of the pattern conservation. Of particular interest to us is the nearest neighbor simple exclusion process in one dimension, for which the long time tracer distribution function is known to be~\cite{harris1965_diffusion,levitt1973_dynamics,alexander1978_tracer,vanBeijeren1983_diffusion}:
\begin{equation} \label{eq:3.1.1}
G_{\mathrm{tr}}(\ell,t)\rightarrow G^{(nI)}_{\mathrm{tr}}(\ell,t) := \frac{1}{(16Dt\pi^2)^{1/4}}\exp\Bigl\{ - \frac{\ell^2}{4\sqrt{Dt}} \Bigr\},
\end{equation}
where the superscript $(nI)$ indicates that we are considering generic, \textit{non-integrable} systems. $G^{(nI)}_{\mathrm{tr}}(\ell,t)$ takes the form of a Gaussian that broadens subdiffusively slowly,
\begin{equation} \label{eq:3.1.2}
\begin{split}
\braket{\Delta \ell(t)^2} = 2\sqrt{Dt}.
\end{split}
\end{equation}
The generalized diffusion constant $D$ is determined via the density $\rho$ of particles on the chain and the bare hopping rate $\Gamma$ per time step of an inividual particle,
\begin{equation} \label{eq:3.1.3}
D = \frac{\Gamma}{\pi}\bigl(\rho^{-1}-1\bigr)^2.
\end{equation}
We consider two examples in detail in the following, the random $tJ_z$ -- model and the random XNOR model, for which \eqs{eq:3.1.1}{eq:3.1.3} will provide us with the exact long time spin correlations after identifying conserved spin patterns in the appropriate variables.

\subsection{Random circuit $tJ_z$ -- model}

\begin{figure}[t]
\centering
\includegraphics[trim={0cm 0cm 0cm 0cm},clip,width=0.99\linewidth]{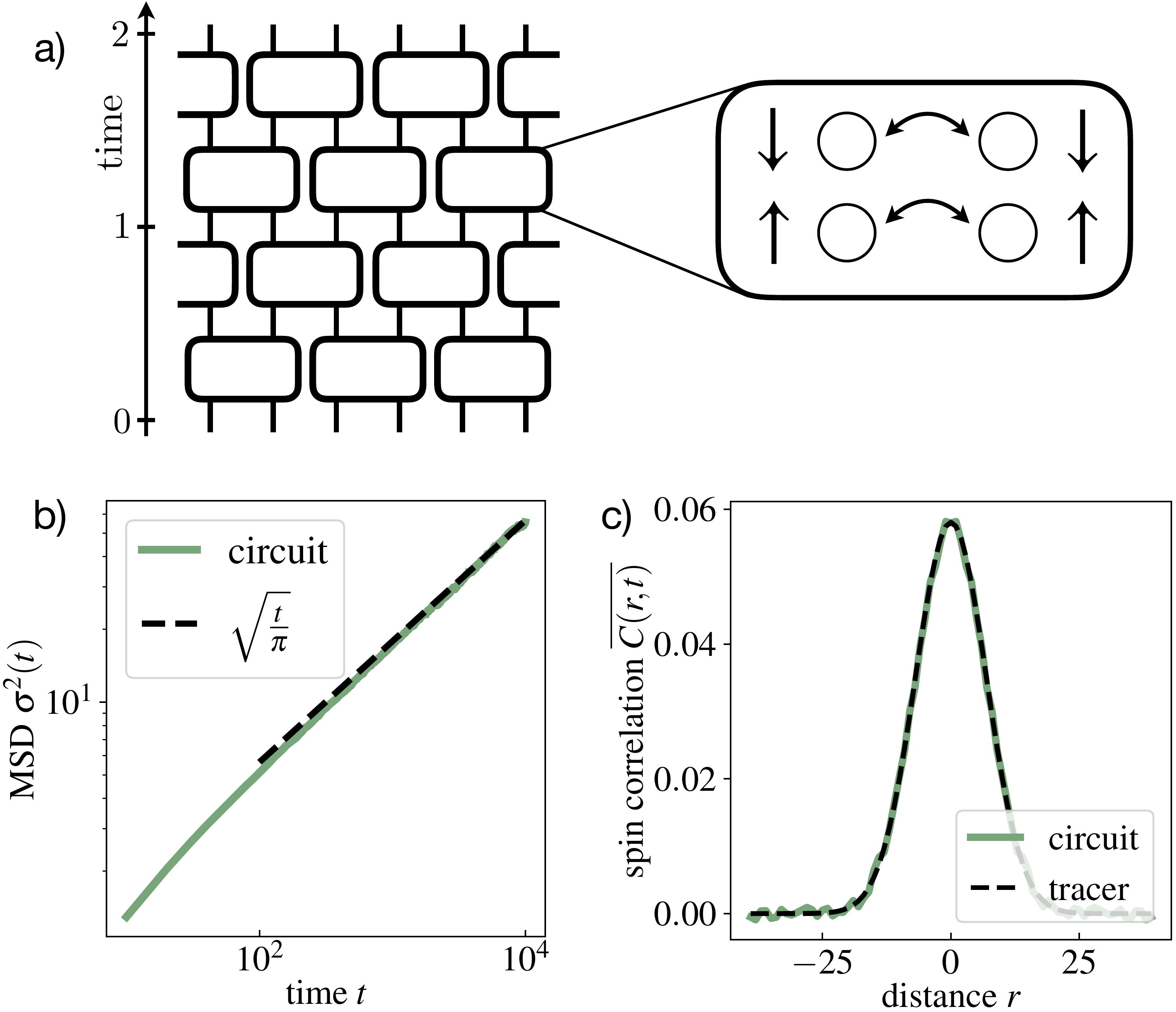}
\caption{\textbf{Random $tJ_z$ -- circuit.} \textbf{(a)} In the random $tJ_z$ -- circuit we consider a three-state local Hilbert space and two-site local random unitary gates. The local gates connect states with spins hopping between neighboring lattice sites. \textbf{(b)} The mean squared displacement (MSD) associated to the spin correlations $C(r,t)$ agrees with the tracer prediction of \eq{eq:3.2.5} for particle density $\rho=2/3$ and single particle hopping rate $\Gamma = 1$. \textbf{(c)} The form of the correlations $C(r,t)$ at time $t=7000$ of the circuit evolution. It assumes a Gaussian shape as expected from the tracer distribution. We obtain these numerical results by sampling the discrete stochastic Markov process of \eq{eq:3.2.3}. The data was averaged over $1000$ randomly chosen product initial states of the Markov process in a system of length $L=5000$.}
\label{fig:2}
\end{figure}

Our first example is a direct implementation of a random version of the $tJ_z$ -- model. We consider a chain with local Hilbert space spanned by the states $\ket{q}=\ket{-1}$, $\ket{0}$, $\ket{1}$. Each basis state can be written as $\ket{\bs{q}}=\ket{\bs{x},\bs{\sigma}}$ and we demand that the pattern $\bs{\sigma}$ of $\pm 1$-spins be a constant of motion. We then consider a random unitary time evolution given by
\begin{equation} \label{eq:3.2.1}
\hat{U}(t) = \prod_{\ell=1}^{tL} \hat{U}_{\ell},
\end{equation}
where individual two-site gates $\hat{U}_\ell$ are arranged spatially as shown in \fig{fig:2}. Each of the $\hat{U}_{\ell}$ is given by
\begin{equation} \label{eq:3.2.2}
\hat{U}_{\ell} = \sum_{s} \hat{P}_s \hat{U}_s \hat{P}_s,
\end{equation}
where $s$ labels the symmetry sectors of the two-site local Hilbert space that are connected under the constraint of keeping $\bs{\sigma}$ constant. Specifically, there are five sectors that contain only a single local configuration, $\{\ket{-1 -1}\}$, $\{\ket{-1 1}\}$, $\{\ket{1 -1}\}$, $\{\ket{1 1}\}$, and $\{\ket{0 0}\}$, as well as two sectors that contain two states each, $\{\ket{0,1},\ket{1,0}\}$, $\{\ket{0,-1},\ket{-1,0}\}$. $\hat{P}_s$ is a projector onto these connected sectors. The unitary operators $\hat{U}_s$ acting within each sector are then chosen randomly from the Haar measure.

Averaging the time evolution over the random gates, the associated circuit-averaged probabilities required to compute the spin correlations $\overline{C(r,t)}$ are given by a classical discrete Markov process. Specifically, we follow Ref.~\cite{Singh2021_subdiff} in introducing the notation $|\bs{x},\bs{\sigma}) := \ket{\bs{x},\bs{\sigma}}\bra{\bs{x},\bs{\sigma}}$ for the projector onto the state $\ket{\bs{x},\bs{\sigma}}$, as well as an associated inner product $(\hat{A}|\hat{B}) := \mathrm{Tr}\bigl[ \hat{A}\hat{B}^\dagger \bigr]$ for operators. The matrix elements for the time evolution are then given by~\cite{Singh2021_subdiff}
\begin{equation} \label{eq:3.2.3}
\overline{|a(\bs{x}^\prime\bs{\sigma}|\bs{x}\bs{\sigma};t)|^2} = (\bs{x}^\prime,\bs{\sigma}|\hat{\mathcal{T}}^t|\bs{x},\bs{\sigma}),
\end{equation}
with a transfer matrix $\hat{\mathcal{T}}$ given by
\begin{equation} \label{eq:3.2.4}
\begin{split}
\hat{\mathcal{T}} &= \bigotimes_{\ell=1}^{L} \hat{\mathcal{T}}_\ell \\
\hat{\mathcal{T}}_\ell &= \sum_s \frac{1}{d_s} \sum_{s_1,s_2 \in s} |s_1)(s_2|,
\end{split}
\end{equation}
where $d_s$ is the size of the local two-site symmetry sector $s$.
We see that \eqs{eq:3.2.3}{eq:3.2.4} describe the averaged probabilities $\overline{|a(\bs{x}^\prime\bs{\sigma}|\bs{x}\bs{\sigma};t)|^2}$ in terms of a stochastic lattice gas: For each applied gate a particle hops with probability $1/2$ to an empty neighboring site and stays at its position if the neighboring site is occupied by another particle. In particular, $\overline{|a(\bs{x}^\prime\bs{\sigma}|\bs{x}\bs{\sigma};t)|^2} = \overline{|a(\bs{x}^\prime|\bs{x};t)|^2}$ is \textit{independent} of the spin pattern $\bs{\sigma}$. Thus, the contribution $\overline{R(x,t)}$ of equation \eq{eq:3.0.2} vanishes due to cancellation of positive and negative spin contributions. The long-time mean squared displacement of the tracer process is in turn exactly described by the simple nearest-neighbor exclusion process through \eqss{eq:3.1.1}{eq:3.1.3}. In our case, the density of particles is given by $\rho=2/3$ at infinite temperature. Furthermore, for a single time step consisting of two layers as shown in \figc{fig:2}{a} there are two attempted moves at rate $1/2$ per particle, such that we can effectively set $\Gamma = 1$. 
This yields $D=1/4\pi$ for the random circuit $tJ_z$ --  model, see also \fig{fig:1}. The theory of tracer diffusion of hard core particles in one dimension thus predicts a mean squared displacement 
\begin{equation} \label{eq:3.2.5}
\sigma^2(t):= \sum_r r^2 \, \overline{C(r,t)} = \sqrt{t/\pi}
\end{equation}
at long times $t$ with a Gaussian shape of the averaged correlations $\overline{C(r,t)}$. The mean squared displacement thus grows subdiffusively $\sim \sqrt{t}$ as opposed to conventional diffusive growth $\sim t$. We confirm this prediction by numerically sampling the stochastic Markov process \eq{eq:3.2.3} which yields the spin correlations in \figc{fig:2}{b+c}.

\begin{figure}[t]
\centering
\includegraphics[trim={0cm 0cm 0cm 0cm},clip,width=0.99\linewidth]{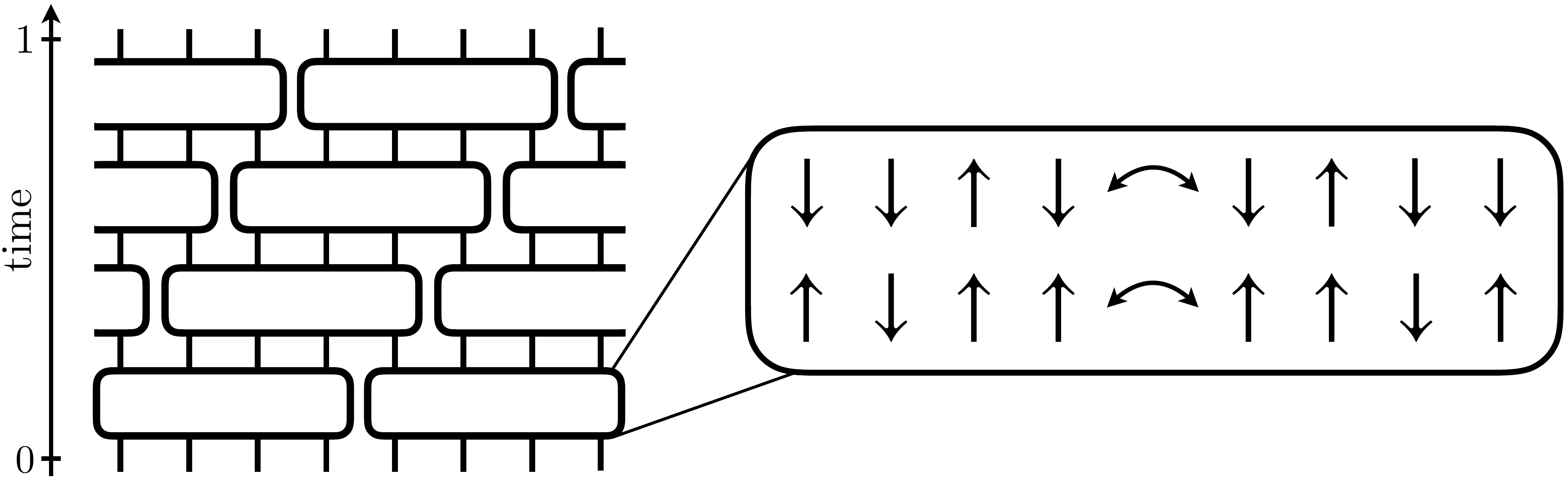}
\caption{\textbf{Random XNOR circuit.} In the random XNOR circuit we consider a two-state local Hilbert space and four-site local random unitary gates. The allowed moves under these gates connect local configurations by exchanging nearest neighbor states, conditioned on the two surrounding sites being in the same state.}
\label{fig:3}
\end{figure}

\subsection{Random circuit XNOR model}
We consider a second example of generic unitary quantum dynamics where we can use the tracer formulae \eqss{eq:3.1.1}{eq:3.1.3} to derive the long-time behavior of local spin correlations, the random XNOR circuit~\cite{Singh2021_subdiff}.
The model is an effective spin $S=1/2$ system with Hilbert space spanned by the local states $\ket{\uparrow},\ket{\downarrow}$. The local unitaries $\hat{U}_{\ell}$ that generate the time evolution are \textit{four-site} gates that conserve both the total magnetization $\hat{M} = \sum_x \hat{S}^z_x$ as well as the number of Ising domain walls $\hat{D}=\sum_x (\hat{S}^z_{x+1}-\hat{S}^z_x)^2$. Therefore, $\hat{U}_{\ell}$ can exchange the central two spins only if the outer two spins have the same value, see \fig{fig:3}. Writing $\hat{U}_{\ell} = \sum_s \hat{P}_s \hat{U}_s \hat{P}_s$ as in \eq{eq:3.2.2}, the only symmetry sectors $s$ that contain more than a single state are $\{\ket{\uparrow,\uparrow,\downarrow,\uparrow},\ket{\uparrow,\downarrow,\uparrow,\uparrow}\}$ and $\{\ket{\downarrow,\uparrow,\downarrow,\downarrow},\ket{\downarrow,\downarrow,\uparrow,\downarrow}\}$.
We refer to this system as the random XNOR model following Ref.~\cite{Singh2021_subdiff}, which established that spin correlations in this model show subdiffusive transport with $z=4$. While the dynamical exponent is in agreement with the tracer picture, the $tJ_z$ -- like existence of a conserved pattern is not immediately apparent in the random XNOR model. We first describe the mapping to such a conserved pattern using the original spin variables $S^z_x \in \{\uparrow,\downarrow\}$, see also Refs.~\cite{dias2000_exact,Tomasi19}. We then construct an equivalent mapping using domain wall variables $\hat{S}^z_{x+1}-\hat{S}^z_x$, see also Refs.~\cite{menon1997_dimers,yang2020_strict,pozsgay2021_fxxz}. Combining both pictures, we will be able to explain the full form of the spin correlations $\overline{C(r,t)}$ at long times.

\begin{figure}[t]
\centering
\includegraphics[trim={0cm 0cm 0cm 0cm},clip,width=0.8\linewidth]{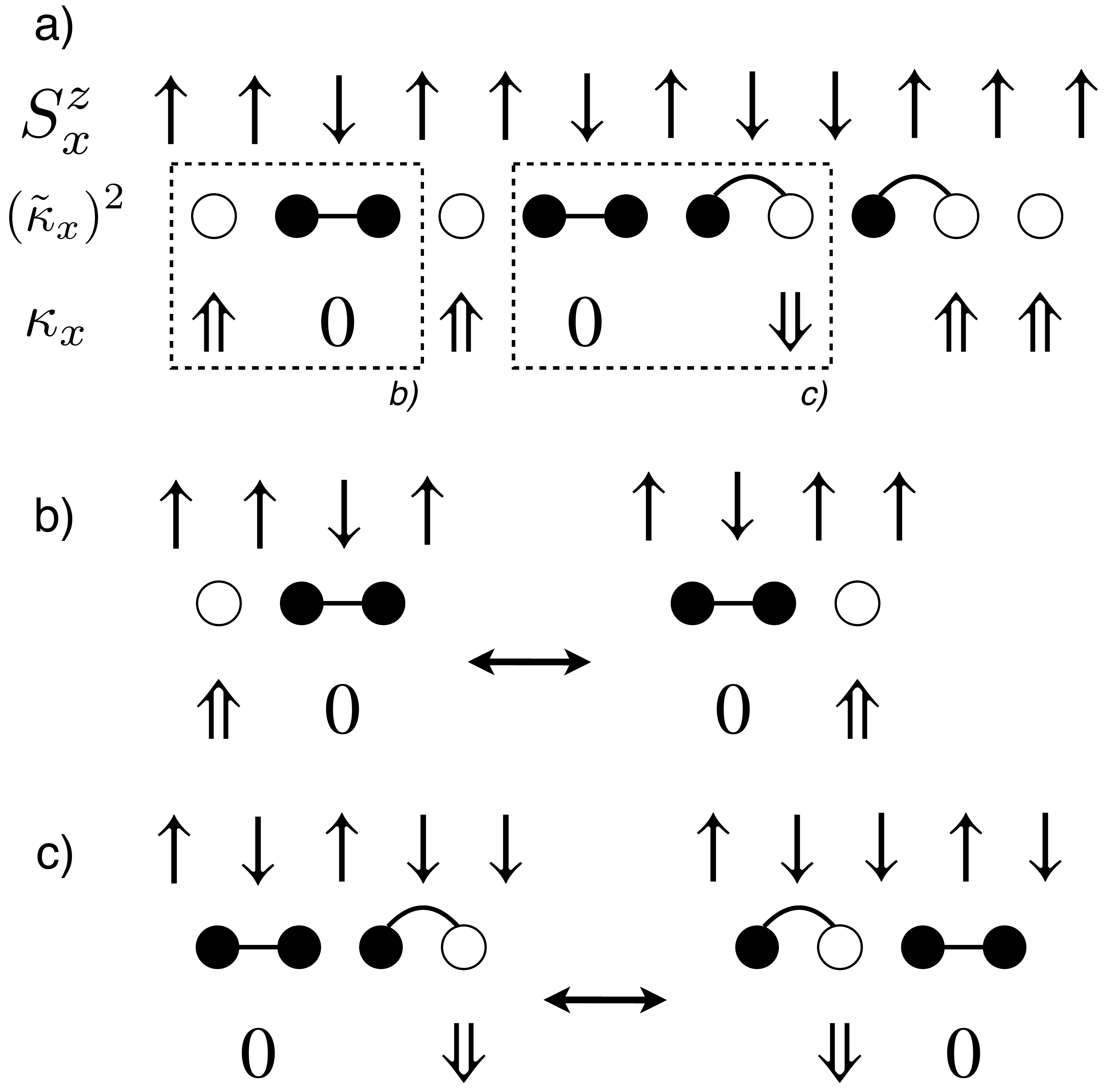}
\caption[\textbf{Conserved charge pattern in the random XNOR circuit.}]{\textbf{Conserved charge pattern in the random XNOR circuit.} \textbf{a)} Mapping between original spin-$1/2$ degrees of freedom and an effective conserved superspin pattern. Going from left to right, two neighboring aligned spins are mapped to a corresponding superspin $\Uparrow$ or $\Downarrow$, while two neighboring domain walls map to a vacant site $0$. The pattern of non-zero superspins is conserved under random XNOR dynamics. In the domain wall picture (empty and filled circles), the mobile vacancies correspond to domain wall pairs (\dwpair). Furthermore, in the domain wall picture the objects that form a conserved pattern are given by single bonds of aligned spins (\alsingle) as well as single domain walls paired up with a neighboring aligned bond (\dwalpair). This  domain wall conserved pattern is obtained by removing all mobile domain wall pairs (\dwpair) see also \fig{fig:5}. \textbf{b)+c)} Elementary XNOR moves within the above mapping. In superspin language, vacancies $0$ and superspins $\Uparrow/\Downarrow$ exchange positions. In domain wall language, a mobile domain wall pair exchanges positions with one of the objects contributing to the domain wall pattern (i.e. \alsingle or \dwalpair).}
\label{fig:4}
\end{figure}

\subsubsection{Conserved pattern: spin picture}
Let us consider a state $\ket{\bs{s}}$ with $s_x \in \{\uparrow,\downarrow\}$ in a system of length $L$. We map this state (bijectively up to boundary terms) to a state $\ket{\bs{\tau}}$ of effective `superspin' degrees of freedom $\tau_i \in \{\Uparrow, 0 , \Downarrow\}$ on a chain of length $\tilde{L}(\bs{s})$ which explicitly depends on the state $\ket{\bs{s}}$ in the original spin-$1/2$ picture, see also Refs.~\cite{dias2000_exact,Tomasi19}. We start at the left end $x=-L/2$ of the original chain and consider the bond between the first two spins $s_{-L/2},s_{-L/2+1}$. There are two possibilities: 
\begin{enumerate}
\item If $s_{-L/2}=s_{-L/2+1}$, we add $\tau = \Uparrow$ and $\tau=\Downarrow$ to the superspin configuration, for $s_{-L/2}=\uparrow$ and $s_{-L/2}=\downarrow$, respectively.
\item If $s_{-L/2}\neq s_{-L/2+1}$, consider the next bond between $s_{-L/2+1},s_{-L/2+2}$: If $s_{-L/2+1}=s_{-L/2+2}$, we again accordingly add $\tau = \Uparrow$ or $\tau = \Downarrow$ to the superspin configuration for $s_{-L/2+1}=\uparrow$ and $s_{-L/2+1}=\downarrow$, respectively. On the other hand, if also $s_{-L/2+1} \neq s_{-L/2+2}$, we add the superspin $\tau = 0$.
\end{enumerate}
The above steps determine the first element (from the left) of the superspin configuration. The next superspin is determined by moving to the next bond between two spins and repeating the above steps. This process is reiterated until all bonds in the original picture have been accounted for, yielding $\ket{\bs{\tau}}=\ket{\bs{\tau}(\bs{s})}$. An example of this mapping is illustrated in \figc{fig:4}{a}.

With the superspin description $\ket{\bs{\tau}}$ we are back to a $tJ$ -- like Hilbert space structure. Under the random XNOR dynamics described above (see also \fig{fig:4}) the number and the pattern of non-zero superspins are indeed conserved: On the one hand, every bond of aligned spins contributes a non-zero superspin and the total number of aligned nearest neighbor spins is constant due to domain wall conservation. On the other hand, two opposite superspins located next to each other, e.g. $\ket{...\Uparrow \Downarrow ...}$, translate into a local configuration of four spins, $\ket{...\uparrow\uparrow\downarrow\downarrow ...}$, on which the XNOR gates of \fig{fig:3} can only act trivially. Hence, $\Uparrow$ and $\Downarrow$ can never exchange relative positions. If we write $\ket{\bs{\tau}}=\ket{\bs{x},\bs{\sigma}}$ as before, the pattern $\bs{\sigma}$ is conserved. Through this mapping we are led back to the random circuit $tJ_z$ -- constraints considered in the previous section, accounting for the dynamical exponent $z=4$. 

In addition, we also analyze in the following how the conserved superspin pattern translates \textit{quantitatively} into the correlations of the original spin variables. To this end, we define the quantities
\begin{equation} \label{eq:3.3.1}
\hat{\kappa}_x := \frac{1}{2} ( \hat{S}^z_x + \hat{S}^z_{x+1} )
\end{equation}
within the spin-$1/2$ picture, which detect whether the bond between the spins at $x,x+1$ contributes a non-zero superspin to $\ket{\bs{\tau}}$. The dynamic correlation function
$\overline{\braket{\hat{\kappa}_{2r}(t)\hat{\kappa}_0(0)}}$
then probes how a non-zero superspin excitation initially located between sites $0,1$ spreads to the bond between $2r,2r+1$. Crucially, according to the random XNOR gates depicted in \fig{fig:4}, the non-zero superspins $\Uparrow,\Downarrow$ \textit{only move by steps of length two} with respect to the original lattice. At the same time they move only by a distance $r$ within the compressed superspin pattern $\bs{\tau}$. Since the superspin description reduces to the random $tJ_z$ -- model analyzed above we can write
\begin{equation} \label{eq:3.3.2}
\overline{\braket{\hat{\kappa}_{2r}(t)\hat{\kappa}_0(0)}} = \frac{1}{2} G_{\mathrm{tr}}^{(nI)}(r,t).
\end{equation}
Here, the prefactor $1/2$ corresponds to the probability of finding a non-zero superspin between sites $0,1$. The hopping rate of superspins entering \eq{eq:3.3.2} through \eq{eq:3.1.3} can again effectively be set to $\Gamma=1$ for the circuit geometry of \fig{fig:3}.
On the other hand, care needs to be taken to determine the density $\rho$ of non-zero superspins, as the infinite temperature average in the original spin variables $s_x$ \textit{does not} transfer directly to an infinite temperature average in the superspin picture. We will derive this density below in the domain wall picture, see \eq{eq:3.3.14}; for now we quote only the obtained result $\rho = 3/4$ entering \eq{eq:3.3.2}.

Inserting the definition \eq{eq:3.3.1} of $\hat{\kappa}_x$ back into \eq{eq:3.3.2} we obtain
\begin{equation} \label{eq:3.3.3}
\begin{split}
& \frac{1}{2} G_{\mathrm{tr}}^{(nI)}(r,t) = \frac{1}{4}\Bigl( \overline{\braket{\hat{S}^z_{2r-1}(t)\hat{S}^z_{0}(0)}} + \\
& \qquad \qquad + 2\overline{\braket{\hat{S}^z_{2r}(t)\hat{S}^z_{0}(0)}} + \overline{\braket{\hat{S}^z_{2r+1}(t)\hat{S}^z_{0}(0)}} \Bigr),
\end{split}
\end{equation}
where we made use of translational invariance in the bulk. We could have performed an equivalent calculation for the correlation $\overline{\braket{\hat{\kappa}_{2r+1}(t)\hat{\kappa}_0(0)}}$ and thus
\begin{equation} \label{eq:3.3.4}
\begin{split}
& \frac{1}{2} G_{\mathrm{tr}}^{(nI)}(r,t) = \frac{1}{4}\Bigl( \overline{\braket{\hat{S}^z_{2r}(t)\hat{S}^z_{0}(0)}} + \\
& \qquad \qquad + 2\overline{\braket{\hat{S}^z_{2r+1}(t)\hat{S}^z_{0}(0)}} + \overline{\braket{\hat{S}^z_{2r+2}(t)\hat{S}^z_{0}(0)}} \Bigr),
\end{split}
\end{equation}
which will be relevant for resolving the $A/B$-sublattice structure further below.
From \eq{eq:3.3.3} we then obtain the mean squared displacement at long times,
\begin{equation} \label{eq:3.3.5}
\begin{split}
&\sigma^2(t) = \sum_r r^2 \overline{\braket{\hat{S}^z_{r}(t)\hat{S}^z_{0}(0)}} \\
& \xrightarrow{t \gg 1} \frac{1}{2} \sum_r (2r)^2 \Bigl\{ \overline{\braket{\hat{S}^z_{2r-1}(t)\hat{S}^z_{0}(0)}} + \\
& \qquad\qquad + 2 \overline{\braket{\hat{S}^z_{2r}(t)\hat{S}^z_{0}(0)}} + \overline{\braket{\hat{S}^z_{2r+1}(t)\hat{S}^z_{0}(0)}} \Bigr\} = \\ 
&= \sum_r (2r)^2 G_{\mathrm{tr}}^{(nI)}(r,t) = \sum_r r^2 G_{\mathrm{tr}}^{(nI)}(r,16t) = \frac{8}{3\sqrt{\pi}} \sqrt{t},
\end{split}
\end{equation}
where we used \eqs{eq:3.1.2}{eq:3.1.3} with $\rho=3/4$ and $\Gamma=1$. We can absorb the factor of $16$ in $G_{\mathrm{tr}}^{(nI)}(r,16t)$ into the effective (sub)diffusion constant to obtain $D=16/9\pi$, see also \fig{fig:1}. Again the dynamics is subdiffusive with $z=4$. 

To verify this prediction we simulate the circuit of \fig{fig:3} numerically, again using the mapping to a classical Markov process. The derivation of the associated transfer matrix $\hat{\mathcal{T}}$ proceeds in full analogy to the $tJ_z$ -- case. \figc{fig:6}{b} demonstrates the validity of \eq{eq:3.3.5}. In addition, \eq{eq:3.3.3} predicts a Gaussian enveloping shape of the charge correlations, which we numerically verify in \figc{fig:6}{a}. Intriguingly however, \eq{eq:3.3.3} in principle allows for additional sublattice structure. We indeed find sizeable staggered oscillations on top of the Gaussian in \figc{fig:6}{a}. These oscillations do not decay at large times and thus hint at additional structure in the model. In order to explain and quantitatively describe these short-distance oscillations, we will switch to a domain wall picture in the following.

\subsubsection{Conserved pattern: domain wall picture}

\begin{figure}[t]
\centering
\includegraphics[trim={0cm 0cm 0cm 0cm},clip,width=0.99\linewidth]{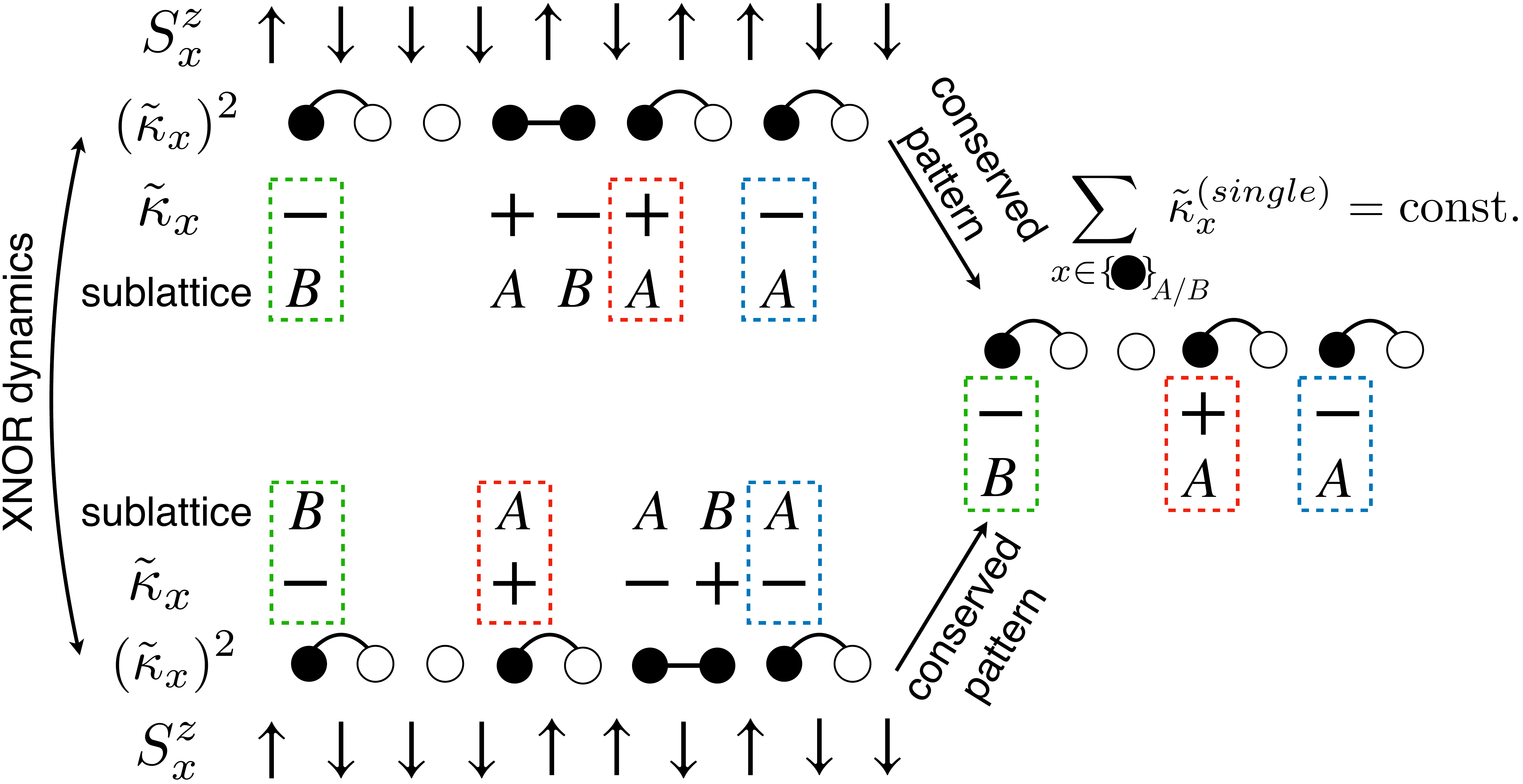}
\caption[\textbf{Sublattice symmetry.}]{\textbf{Sublattice symmetry.} In the domain wall picture, we obtain a conserved pattern by removing all mobile domain wall pairs (illustrated as \dwpair). The resulting pattern is formed by single aligned bonds (\alsingle) and pairs of a single domain wall with an aligned bond to its right (\dwalpair). The pattern is subject to an exclusion constraint of nearest neighbor domain walls, i.e. every filled circle necessarily has at least two empty circle neighbors (one to the left and one to the right). The sublattice charge of all charged domain walls $\hat{\tilde{\kappa}}$ that are not part of a mobile pair is then conserved. This is due to the mobile pairs having a spatial extension of length two, such that single domain walls exchanging their position with such pairs only move by steps of length two at a time, hence preserving their sublattice. They also preserve their charge value since the domain wall charges are perfectly anticorrelated globally.}
\label{fig:5}
\end{figure}

An alternative description of the Hilbert space structure can be given in terms of the domain wall charge variables
\begin{equation} \label{eq:3.3.6}
\hat{\tilde{\kappa}}_x := \frac{1}{2}(\hat{S}^z_{x+1}-\hat{S}_x^z),
\end{equation}
which ascribes a sign depending on whether the local configuration is $\ket{...\downarrow\uparrow...}$ or $\ket{...\uparrow\downarrow...}$. After fixing the leftmost spin, a complete description of a spin configuration is also given in terms of the locations of its domain walls, $(\tilde{\kappa}_x)^2$, regardless of their sign. We can construct a domain wall version of the conserved charge pattern, see also Refs.~\cite{menon1997_dimers,yang2020_strict,pozsgay2021_fxxz}: Starting from the left of the system, neighboring domain walls are paired up into mobile pairs \dwpair, see \figc{fig:4}{a}. The remaining single domain walls are paired up with their corresponding right neighbor bond that connects two aligned spins, \dwalpair. Defined in this way, the dynamics in the system is generated by mobile domain wall pairs (\dwpair) moving through the system, exchanging positions with single bonds of aligned spins (\alsingle) and with the pairs of domain walls and aligned bonds (\dwalpair). The elementary dynamical processes are depicted in \figc{fig:4}{b+c}. By removing all mobile domain wall pairs, see e.g. \fig{fig:5}, we obtain a conserved pattern in the domain wall description formed by single aligned bonds \alsingle and the pairs \dwalpair of domain wall and aligned bond. The conserved pattern thus exhibits a blockade of nearest neighbor domain walls. The number of conserved patterns with such a blockade grows as a Fibonacci sequence in system size. This Fibonacci number then also corresponds to the number of disconnected subsectors in the Hilbert space~\cite{Tomasi19,yang2020_strict}.

\begin{figure*}[t]
\centering
\includegraphics[trim={0cm 0cm 0cm 0cm},clip,width=0.99\linewidth]{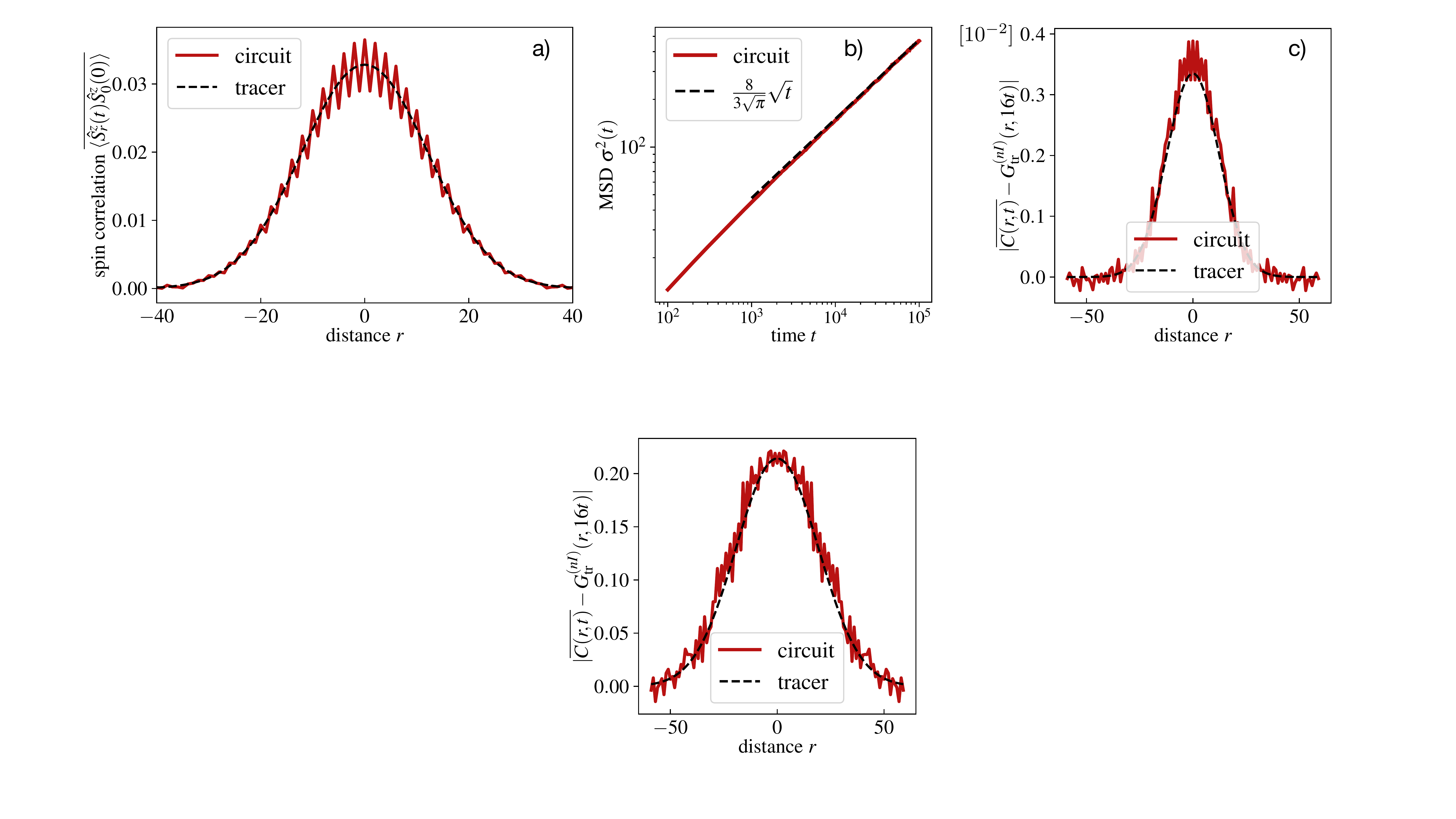}
\caption{\textbf{Random XNOR model: Numerical results.} \textbf{a)} Spatial shape of the spin correlation function at time $t=10^4$ of the random circuit shown in \fig{fig:3}. The black dashed line corresponds to the Gaussian of the tracer probability distribution predicted by \eq{eq:3.3.19}. \textbf{b)} The mean squared displacement $\sigma^2(t)$ of the spin correlations agrees with the late time prediction of \eq{eq:3.3.5}. \textbf{c)} The absolute value $\bigl|C(r,t)-G_{\mathrm{tr}}^{(nI)}(x,16t)\bigr|$ of the oscillations on top of the Gaussian shape as seen in a). The black dashed line corresponds to the prediction of \eq{eq:3.3.19}. The results demonstrate the quantitative accuracy of the tracer description. The data was averaged over $12800$ random initial states of the associated stochastic Markov process in a system of length $L=5000$.}
\label{fig:6}
\end{figure*}

Due to the two-site spatial extension of the domain wall pairs, the total $A/B$ sublattice charge of all domain walls $\hat{\tilde{\kappa}}_x$ that are not part of a mobile pair is a conserved quantity, see \fig{fig:5}. Formally, we can express
\begin{equation} \label{eq:3.3.7}
\hat{\tilde{\kappa}}_x=\hat{\tilde{\kappa}}^{(single)}_x+\hat{\tilde{\kappa}}^{(pair)}_x,
\end{equation} 
where $\hat{\tilde{\kappa}}^{(single)}_x$ is the domain wall charge operator for \textit{single} domain walls while $\hat{\tilde{\kappa}}^{(pair)}_x$ is the operator for domain walls that are part of mobile \textit{pairs}. The two sublattice charges 
\begin{equation} \label{eq:3.3.8}
\hat{\mathcal{Q}}_{A/B} = \sum_{x \in A/B} \hat{\tilde{\kappa}}^{(single)}_x = \mathrm{const.}
\end{equation}
are conserved quantities. The operator $\hat{\tilde{\kappa}}_x$ thus separates into a part that has overlap with the sublattice conservation laws of \eq{eq:3.3.8} and a part not associated to any such conservation law ($\sum_{x \in A/B} \hat{\tilde{\kappa}}^{(pair)}_x \neq \mathrm{const.}$).
The domain wall charge correlation function on the even sublattice at late times will thus be dominated by the transport associated to the conserved quantity $\hat{\mathcal{Q}}_A$, i.e.
\begin{equation} \label{eq:3.3.9}
\overline{\braket{\hat{\tilde{\kappa}}_{2r}(t)\hat{\tilde{\kappa}}_0(0)}} \stackrel{t \gg 1}{=} \overline{\braket{\hat{\tilde{\kappa}}^{(single)}_{2r}(t)\hat{\tilde{\kappa}}^{(single)}_0(0)}}.
\end{equation}
Corrections to \eq{eq:3.3.9} are expected to decay quickly (generically exponentially fast) in time. Since the single domain wall charges are part of a conserved pattern as described above their dynamical correlations are again given by the tracer distribution,
\begin{equation} \label{eq:3.3.10}
\overline{\braket{\hat{\tilde{\kappa}}_{2r}(t)\hat{\tilde{\kappa}}_0(0)}} \stackrel{t \gg 1}{=} C_0 \, G_{\mathrm{tr}}^{(nI)}(r,t),
\end{equation}
with constant prefactor $C_0$ to be determined. The density $\rho$ and the hopping rate $\Gamma$ entering \eq{eq:3.3.10} are the same as in the spin picture above. 

To compute $C_0$ we equate \eqs{eq:3.3.9}{eq:3.3.10} and take the sum over $r$,
\begin{equation} \label{eq:3.3.11}
C_0 = \sum_r \braket{\hat{\tilde{\kappa}}^{(single)}_{2r}\hat{\tilde{\kappa}}^{(single)}_0},
\end{equation}
resulting in a simple \textit{static} (notice the absence of the circuit average) correlation function at infinite temperature. We can express \eq{eq:3.3.11} as
\begin{equation} \label{eq:3.3.12}
C_0 = \Braket{\bigl(\hat{\tilde{\kappa}}^{(single)}_0\bigr)^2} \sum_r \Braket{\hat{\tilde{\kappa}}^{(single)}_{2r}}^\prime,
\end{equation}
where $\braket{\cdot}^\prime$ denotes a modified infinite temperature average with a single positive domain wall \textit{fixed} to sit at site $0$. The factor $\Braket{\bigl(\hat{\tilde{\kappa}}^{(single)}_0\bigr)^2}$ is determined within the mapping to a conserved pattern from \fig{fig:4}: $\Braket{\bigl(\hat{\tilde{\kappa}}^{(single)}_0\bigr)^2} = \#(\dwalpair)/L$ corresponds to the density of single domain walls paired up with a neighboring aligned spin bond. Using that the overall density of domain walls is $1/2$ and that by symmetry $\#(\dwalpair)/L = \#(\dwpair)/L$, we obtain
\begin{equation} \label{eq:3.3.13}
\begin{split}
\frac{L}{2} &= 2\cdot \#(\dwpair) + \#(\dwalpair) = 3\cdot \#(\dwalpair) \\
&\rightarrow \Braket{\bigl(\hat{\tilde{\kappa}}^{(single)}_0\bigr)^2} = \#(\dwalpair)/L = 1/6.
\end{split}
\end{equation}
With this result we can also compute the density $\rho$ of hard core particles (corresponding to the density of non-zero superspins of the previous section) that we used in \eq{eq:3.3.5}:
\begin{equation} \label{eq:3.3.14}
\rho = 1 - \frac{\#(\dwpair)}{\#(\dwpair)+\#(\alsingle)} = 1 - \frac{L/6}{L/6+L/2} = 3/4.
\end{equation}
The remaining correlation function $\Braket{\hat{\tilde{\kappa}}^{(single)}_{2r}}^\prime$ in \eq{eq:3.3.12} refers only to single domain walls and can thus be computed within the ensemble of possible conserved domain wall patterns, see \fig{fig:5}. Making use of this property we derive the exact value of this correlation function in Appendix~\ref{sec:App1} and quote here our final result
\begin{equation} \label{eq:3.3.15}
C_0 = \frac{1}{12}\bigl(\varphi-1\bigr) 
\end{equation}
for the constant $C_0$, where $\varphi=(1+\sqrt{5})/2$ is the golden ratio.
Then, inserting the definition \eq{eq:3.3.6} in \eq{eq:3.3.10} yields
\begin{equation} \label{eq:3.3.16}
\begin{split}
& C_0\, G_{\mathrm{tr}}^{(nI)}(r,t) = \frac{1}{4}\Bigl( 2\overline{\braket{\hat{S}^z_{2r}(t)\hat{S}^z_{0}(0)}} \\
& \qquad \qquad -\overline{\braket{\hat{S}^z_{2r-1}(t)\hat{S}^z_{0}(0)}} - \overline{\braket{\hat{S}^z_{2r+1}(t)\hat{S}^z_{0}(0)}} \Bigr).
\end{split}
\end{equation}
Performing the equivalent derivation for the correlations $\overline{\braket{\hat{\tilde{\kappa}}_{2r+1}(t)\hat{\tilde{\kappa}}_0(0)}}$ gives us the additional relation
\begin{equation} \label{eq:3.3.17}
\begin{split}
& -C_0\, G_{\mathrm{tr}}^{(nI)}(r,t) = \frac{1}{4}\Bigl( 2\overline{\braket{\hat{S}^z_{2r+1}(t)\hat{S}^z_{0}(0)}} \\
& \qquad \qquad -\overline{\braket{\hat{S}^z_{2r}(t)\hat{S}^z_{0}(0)}} - \overline{\braket{\hat{S}^z_{2r+2}(t)\hat{S}^z_{0}(0)}} \Bigr).
\end{split}
\end{equation}
Adding \eqs{eq:3.3.3}{eq:3.3.16} as well as \eqs{eq:3.3.4}{eq:3.3.17} finally yields the long time correlations
\begin{equation} \label{eq:3.3.18}
\begin{split}
\overline{\braket{\hat{S}^z_{2r}(t)\hat{S}^z_{0}(0)}} = \frac{1}{2}(1+2C_0)\, G_{\mathrm{tr}}^{(nI)}(r,t) \\
\overline{\braket{\hat{S}^z_{2r+1}(t)\hat{S}^z_{0}(0)}} = \frac{1}{2}(1-2C_0)\, G_{\mathrm{tr}}^{(nI)}(r,t),
\end{split}
\end{equation}
which we can rewrite as
\begin{equation} \label{eq:3.3.19}
\overline{C(r,t)} = \overline{\braket{\hat{S}^z_{r}(t)\hat{S}^z_{0}(0)}} = [1+2C_0(-1)^r]\, G_{\mathrm{tr}}^{(nI)}(r,16t).
\end{equation}
\eq{eq:3.3.19} explains the staggered oscillations observed numerically in \figc{fig:6}{a}. To check that our quantitative description (i.e. the constant $C_0$ of \eq{eq:3.3.15}) is accurate, we show in \figc{fig:6}{c} that the quantity $\bigl|\overline{C(r,t)}-G_{\mathrm{tr}}^{(nI)}(x,16t)\bigr|$ agrees with the prediction of \eq{eq:3.3.19}. We emphasize that the staggering of the correlations decribed by \eq{eq:3.3.19} persists up to infinite time.

\subsection{Random circuit dimer model}
We mention only briefly a third example of $tJ_z$ -- like dynamics in a dimer model on a bilayer square lattice geometry as introduced in Ref.~\cite{feldmeier2021_fractondimer}. The system has a short direction along which periodic boundaries are chosen, making the geometry quasi one-dimensional, and the time evolution is generated by local plaquette flips of parallel dimers,
\begin{equation} \label{eq:3.5.1}
\hat{H} = -J\sum_{p} \left(\ket{\plaqv}\bra{\plaqh} + h.c. \right).
\end{equation}
The model is equivalent to a model of closed directed loops and site-local charges on a square lattice cylinder with short circumference. The existence of a conserved charge pattern and its role in inducing a dynamical exponent $z=4$ due to the emergence of a hard core tracer problem has been discussed already in Ref.~\cite{feldmeier2021_fractondimer}. Intuitively, the site-local charges in the model are dimers that go in between the two layers, positive or negative depending on which sublattice they occupy. When a charge is enclosed by a loop, it cannot escape the loop under the dynamics of \eq{eq:3.5.1}. In the presence of a finite density of loops that wind across the circumference of the cylinder (with the same chirality), a conserved pattern is formed by the summing up charges always in between two such loops, see \fig{fig:1} and Ref.~\cite{feldmeier2021_fractondimer} for details. The tracer prediction of $z=4$ and the Gaussian shape of the dynamical charge correlations have been verified numerically in Ref.~\cite{feldmeier2021_fractondimer}. Only the precise values of effective (sub)diffusion constants are unknown since the mapping to a conserved pattern is an effective one. 

\section{Integrable quantum systems with conserved pattern} \label{sec:integrable}

The presence of a conserved charge pattern can also be of use in \textit{integrable} $tJ_z$ -- like quantum systems and, in certain cases, provides an alternative route to determine the long time profile and diffusion constant of their spin correlations at infinite temperature. The associated tracer motion relevant for the correlations $C(r,t)$ from \eq{eq:2.10} is one with \textit{ballistically} instead of diffusively moving particles. As single particles move ballistically, the many-body tracer distribution will be diffusive. A similar situation was considered in Refs.~\cite{medenjak2017_diffusion,Klobas2018_exact}, where the spin diffusion constant of an interacting deterministic classical cellular automaton with invariant spin pattern and pattern-independent dynamics was computed exactly. According to \eq{eq:2.10}, this spin diffusion is directly associated with a Gaussian tracer distribution with the same diffusion constant,
\begin{equation} \label{eq:3.4.1}
G_{tr}(\ell,t) \rightarrow G_{tr}^{(I)}(\ell,t) = \frac{1}{\sqrt{4\pi Dt}} \exp\Bigl\{-\frac{\ell^2}{4Dt} \Bigr\}.
\end{equation}
The superscript $(I)$ indicates an integrable process which implies a broadening of $G_{tr}$ as $\sim\sqrt{t}$ (instead of $\sim t^{1/4}$ which we have obtained for generic systems in Sec.~\ref{sec:generic}). Using the result of Ref.~\cite{medenjak2017_diffusion}, the diffusion constant $D$ of \eq{eq:3.4.1} is determined by the density $\rho$ of particles along the chain as well as their effective velocity $v_{\mathrm{eff}}$,
\begin{equation} \label{eq:3.4.2}
\begin{split}
\braket{\Delta x^2} = 2Dt \\
D = \frac{v_{\mathrm{eff}}}{2}(\rho^{-1}-1).
\end{split}
\end{equation}
In the discrete time cellular automaton considered in Ref.~\cite{medenjak2017_diffusion}, all particles have a fixed velocity of $v_{\mathrm{eff}}=2$.

\subsection{Integrable $tJ_z$ -- model}
The above connection can be put to direct use in the integrable $J_z\rightarrow 0$ limit of the $tJ_z$ -- model~\cite{KOTRLA90},
\begin{equation} \label{eq:3.4.3}
\hat{H}_{t} = -\sum_{x,\sigma} ( \hat{\tilde{c}}^\dagger_{x+1,\sigma}\hat{\tilde{c}}^{}_{x,\sigma} + h.c. ),
\end{equation}
which features only the hopping of spinful fermions with forbidden double occupancy. The matrix elements $a(\bs{x}^\prime\bs{\sigma}|\bs{x}\bs{\sigma},t) = a(\bs{x}^\prime|\bs{x},t)$ of the time evolution obtained from \eq{eq:3.4.3} do not depend on the pattern $\bs{\sigma}$, so $R(r,t)=0$ in \eq{eq:3.0.2}, and the mapping to a tracer problem is exact at all times. At infinite temperature, the density of particles is $\rho=2/3$ and we can determine the effective velocity $v_{\mathrm{eff}}$ by noting that $\hat{H}_t$ can be mapped to a problem of spinless free fermions~\cite{KOTRLA90},
\begin{equation} \label{eq:3.4.4}
\hat{H}_{t} = -\sum_k \cos(k) \hat{f}^\dagger_k\hat{f}^{}_k.
\end{equation}
We give an intuitive argument for why this is possible: Since the dynamics is oblivious to the conserved spin pattern, for a given inital basis state we can simply \textit{i)} `write down' the invariant spin pattern for bookkeeping purposes, \textit{ii)} remove the fermions' spins, \textit{iii)} perform the time evolution with a Hamiltonian of spinless fermion hopping (of same hopping strength), and then \textit{iv)} reintroduce the pattern afterwards. Steps \textit{i)} and \textit{iv)} in this mapping are of course highly non-local.

At infinite temperature we then expect all momentum modes of \eq{eq:3.4.4} to be occupied with equal probability. While Ref.~\cite{medenjak2017_diffusion} derived \eq{eq:3.4.2} for an automaton in which every particle has the same absolute velocity, here we need to consider a distribution of different momentum modes. We expect that the mean displacement of the equivalent tracer process depends only on the average of the absolute velocity and thus predict the effective velocity for \eq{eq:3.4.2} to be given by the  average absolute group velocity of \eq{eq:3.4.4}:
\begin{equation} \label{eq:3.4.5}
v_{\mathrm{eff}} = \frac{1}{2\pi}\int_{-\pi}^{\pi} dk \, | \partial_k \cos(k) | = \frac{2}{\pi}.
\end{equation}
This leads to the infinite temperature spin diffusion constant
\begin{equation} \label{eq:3.4.6}
D_t = \frac{1}{2\pi}
\end{equation}
for the quantum model $\hat{H}_t$ of \eq{eq:3.4.3}. \eq{eq:3.4.2} also yields the diffusion constant at infinite temperature but with fixed density $\rho \neq 2/3$ or chemical potential $\mu$ such that $\beta\mu = \mathrm{const.}$ as $\beta\rightarrow\infty$.

\begin{figure}[t]
\centering
\includegraphics[trim={0cm 0cm 0cm 0cm},clip,width=0.99\linewidth]{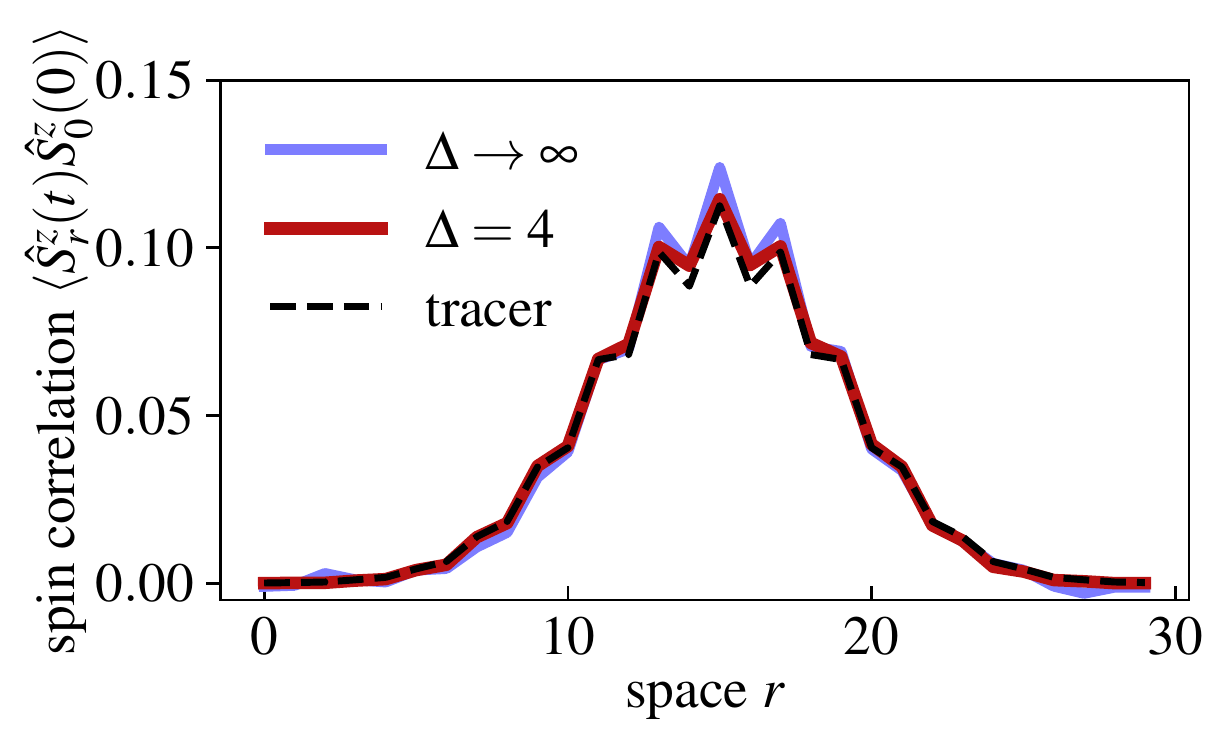}
\caption{\textbf{Folded XXZ chain.} Profile of the dynamical spin correlation function for the folded XXZ chain of \eq{eq:3.4.8} (blue curve, $\Delta \rightarrow \infty$) at time $t=18$ in a system of length $L=30$. The black dashed line shows the prediction of \eq{eq:3.4.10}, there are no fit parameters. The red line shows the profile obtained for the XXZ chain \eq{eq:3.4.7} at anisotropy $\Delta=4$, illustrating good agreement with \eq{eq:3.4.10} at intermediate times, even for moderate values of anisotropy.}
\label{fig:7}
\end{figure}

\subsection{Folded XXZ spin chain}
As a second example we consider an integrable version of the random XNOR model, the folded XXZ chain. It is obtained from the integrable XXZ model
\begin{equation} \label{eq:3.4.7}
\hat{H}_{\mathrm{XXZ}} = - \sum_i \frac{1}{2}(\hat{S}^+_i\hat{S}^-_{i+1} + \hat{S}^-_i\hat{S}^+_{i+1}) + \Delta \hat{S}^z_i\hat{S}^z_{i+1}
\end{equation}
in the limit of large anisotropy $\Delta \rightarrow \infty$, which using a Schrieffer-Wolff transformation yields
\begin{equation} \label{eq:3.4.8}
\hat{H}_{\mathrm{fXXZ}} = - \sum_i (\frac{1}{4}+\hat{S}^z_{i-1}\hat{S}^z_{i+2})(\hat{S}^+_i\hat{S}^-_{i+1}+\hat{S}^-_i\hat{S}^+_{i+1}).
\end{equation}

$\hat{H}_{\mathrm{fXXZ}}$ remains integrable and many of its thermodynamic and dynamic properties have been analyzed in recent works~\cite{Zadnik2021_fxxz1,Zadnik2021_fxxz2,pozsgay2021_fxxz,
bidzhiev2022_fxxz}. We note that $\hat{H}_{\mathrm{fXXZ}}$ consists of four-site spin- and domain wall-conserving terms and thus features the same effective conserved pattern of superspins as the random XNOR model above. In particular, it can be demonstrated that in the superspin picture, the folded XXZ chain becomes equivalent to the integrable limit of the $tJ_z$ -- model from \eq{eq:3.4.3}, $\hat{H}_{\mathrm{fXXZ}} \rightarrow \hat{H}_t$~\cite{dias2000_exact}. As a consequence, the spin diffusion constant $D_{\mathrm{fXXZ}}$ of the folded XXZ chain can be obtained from the diffusion constant $D$ of superspins determined by the Hamiltonian $\hat{H}_t$ of \eq{eq:3.4.3}. In order to relate the two, we recall that the infinite temperature average in the original spin picture implies a density $\rho=3/4$ of superspins in the associated integrable $tJ_z$ -- model. Furthermore, in the original lattice superspins always move by two sites. The variance $\braket{\Delta x^2}=2D_{\mathrm{fXXZ}}t$ of the original spin correlation profile is thus determined by 
\begin{equation} \label{eq:3.4.9}
\begin{split}
\frac{2D_{\mathrm{fXXZ}}\,t}{4} &= \frac{\braket{\Delta x^2}}{4} = 2 D t \\ \rightarrow D_{\mathrm{fXXZ}} &= \frac{4}{3\pi},
\end{split}
\end{equation}
where we have used that $D=1/3\pi$ for $\rho=3/4$ and $v_{\mathrm{eff}}=2/\pi$ in \eq{eq:3.4.2}. The value of $D_{\mathrm{fXXZ}}$ in \eq{eq:3.4.9} is in agreement with previous analytical results~\cite{gopalakrishnan2019_kinetic}, obtained using generalized hydrodynamics~\cite{castro2016_ghd,bertini2016_ghd}, as well as numerical results for the XXZ chain~\cite{karrasch2014_realtime,karrasch2017_hubbard}. In addition, following the analysis of the sublattice domain wall charge in the XNOR model, see \fig{fig:5}, we predict the full long time profile of the dynamical spin correlations for the folded XXZ chain to be
\begin{equation} \label{eq:3.4.10}
C(r,t) = [1+2C_0(-1)^r]\,G_{tr}^{(I)}(r,t).
\end{equation}
Here, $G_{tr}^{(I)}(r,t)$ is from \eq{eq:3.4.1} with $D=D_{\mathrm{fXXZ}}$. In particular, the long time profile features characteristic staggered oscillations of strength $2C_0$ with $C_0$ from \eq{eq:3.3.15}. 

These oscillations also lead to a distinct contribution to the spin conductivity $\sigma(k,\omega)=\frac{\beta}{2}\braket{j(k,\omega)j(-k,-\omega)}$, with spin current $j(k,\omega)$. Using the continuity equation $\partial_t S^z_x(t) + \partial_x j(x,t)= 0$, we relate $\sigma(k,\omega)=\frac{\beta\omega^2}{2k^2}C(k,\omega)$ and obtain
\begin{equation} \label{eq:3.4.11}
\sigma(k=\pi,\omega\rightarrow 0) = \frac{C_0\beta}{2} D_{\mathrm{fXXZ}}.
\end{equation}
The finite momentum conductivity of the folded XXZ chain thus exhibits a finite contribution at $k=\pi$.

We verify \eq{eq:3.4.10} numerically for the folded XXZ chain of \eq{eq:3.4.8} in systems of finite size and at intermediate time scale in \fig{fig:7}. We consider a chain of size $L=30$ spins by making use of the conserved pattern and employing sparse matrix evolution. The infinite temperature average for the spin correlations was approximated by averaging over $10^5$ randomly chosen product initial states for the time evolution. We find good agreement with \eq{eq:3.4.10} already at a time $t=18$ in \fig{fig:7}.
Furthermore, we note that the anisotropic XXZ chain of \eq{eq:3.4.7} has recently been implemented in quantum simulation experiments~\cite{Jepsen2020_xxz}. In a simple perturbative argument, the Hamiltonian $\hat{H}_{\mathrm{fXXZ}}$ should provide a good estimate for the time evolution of $\hat{H}_{\mathrm{XXZ}}$ up to times $t \sim \Delta^2$ in the anisotropy. Signatures of \eq{eq:3.4.10} should thus be visible at intermediate times already at moderate anisotropy strength. We demonstrate this numerically in \fig{fig:7} by time-evolving an infinite temperature density matrix perturbed by a spin excitation in the center of the chain. We use matrix product state methods with a bond dimension $\chi=1200$ for the time evolution with the anisotropic XXZ Hamiltonian of \eq{eq:3.4.7}~\cite{hauschild2018_tenpy}. We find good agreement with the folded limit and the expression in \eq{eq:3.4.10} already for $\Delta=4$ at a time $t=18$ in \fig{fig:7}.

We conclude this section with the following remark: As discussed previously, the folded XXZ Hamiltonian $\hat{H}_{\mathrm{fXXZ}}$ maps to the integrable limit $\hat{H}_t$ of the $tJ_z$ -- model in the superspin picture. Similarly, one can apply the reverse of this mapping to the $tJ_z$ -- like deterministic cellular automaton studied in Refs.~\cite{medenjak2017_diffusion,Klobas2018_exact}, which is effectively a local automaton for the superspins. Under the reverse mapping we then obtain an automaton for spin-$1/2$ variables that is able to mimic the long-time dynamics of the folded XXZ chain. A different cellular automaton mimicking the folded XXZ dynamics, which exhibits different left- and right-mover velocities, has recently been studied in Ref.~\cite{Pozsgay2021_intaut}. In our prescription outlined above the automaton obtained from Ref.~\cite{medenjak2017_diffusion} by inverting the mapping from superspins to spin-$1/2$ is non-local in the spin-$1/2$ variables and features symmetric left- and right-movers of equal speed instead.

\section{Broken pattern conservation} \label{sec:multipole}
In this section, we return to random unitary circuit models but relax the condition of an exactly conserved spin pattern. In particular, we consider constrained `$tJ_m$ -- like' models in which only a certain number $m$ of moments of the spin pattern remains constant, see \eqs{eq:1.1}{eq:1.3}. Remarkably, breaking the pattern conservation does not immediately imply conventional diffusion but the resulting dynamics sensitively depends on the number of conserved moments. To see this, we note that the pattern-internal spin dynamics in the presence of $m$ conserved multipole moments is governed by the following hydrodynamic equation for the coarse-grained spin density $\braket{\hat{\sigma}_x(t)}$~\cite{gromov2020_fractonhydro,feldmeier2020anomalous},
\begin{equation} \label{eq:4.0.1}
\partial_t \braket{\hat{\sigma}_x} + (-1)^{m} D_s\, \partial_x^{2m+2} \braket{\hat{\sigma}_x} = 0.
\end{equation}
\eq{eq:4.0.1} describes (sub)diffusive dynamics with dynamical exponent $z=2m+2$ and its fundamental solution, which corresponds to the spin part of \eq{eq:2.5} via linear response, reads in momentum space:
\begin{equation} \label{eq:4.0.2}
F(k,t) = \frac{1}{\sqrt{2\pi}} \, \exp(-D_s \, k^{2m+2} t), 
\end{equation}
normalized such that $\int dx\, F(x,t) = 1$. Using \eq{eq:4.0.2} in \eq{eq:2.10} we obtain the spin correlations in momentum space,
\begin{equation} \label{eq:4.0.3}
\begin{split}
&C(k,t) = G_{tr}(k,t)\, F(k,t) = \\
&=\frac{1}{2\pi} \exp\Bigl\{-D_s \, k^{2m+2} t - \sqrt{Dt}\, k^2\Bigr\},
\end{split}
\end{equation}
which we will analyze for different values of $m$ in the following. While $m=0$ leads to conventional diffusion, $m\geq 2$ preserves the tracer mapping at long times. The case $m=1$ turns out to be special, with a competition between two processes that have the same dynamical exponent but different scaling functions.

\begin{figure}[t]
\centering
\includegraphics[trim={0cm 0cm 0cm 0cm},clip,width=0.99\linewidth]{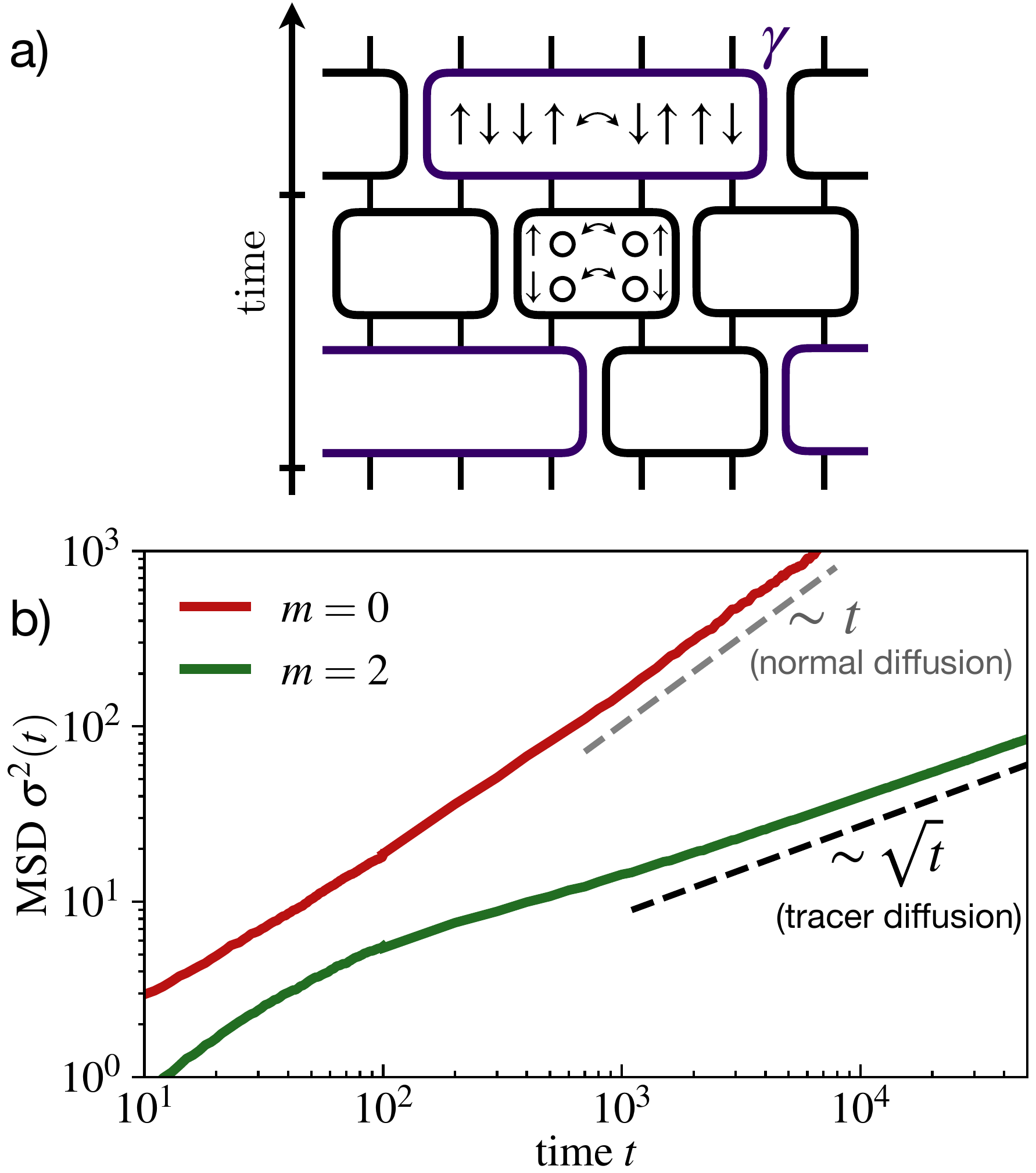}
\caption{\textbf{Moment-conserving random circuit evolution.}
\textbf{a)} We consider $tJ_m$ -- like random circuits with spin terms that conserve a given number $m$ of multipole moments (in the depicted example the dipole moment $m=1$) of the spin pattern. The spin terms are applied with some probability $\gamma$, the particle hopping terms with probability $1-\gamma$.
\textbf{b)} The dynamics of random circuits with multipole moment conserving spin pattern depends on the number $m$ of conserved moments. If only the total spin of the pattern is conserved ($m=0$), the mean-squared displacement $\sigma^2(t)\sim t$ follows conventional diffusion. If all moments up to the  quadrupole moment ($m=2$) or higher are conserved, $\sigma^2(t)\sim \sqrt{t}$ remains dominated by anomalously slow hard core tracer motion.}
\label{fig:8}
\end{figure}

\begin{figure*}[t]
\centering
\includegraphics[trim={0cm 0cm 0cm 0cm},clip,width=0.99\linewidth]{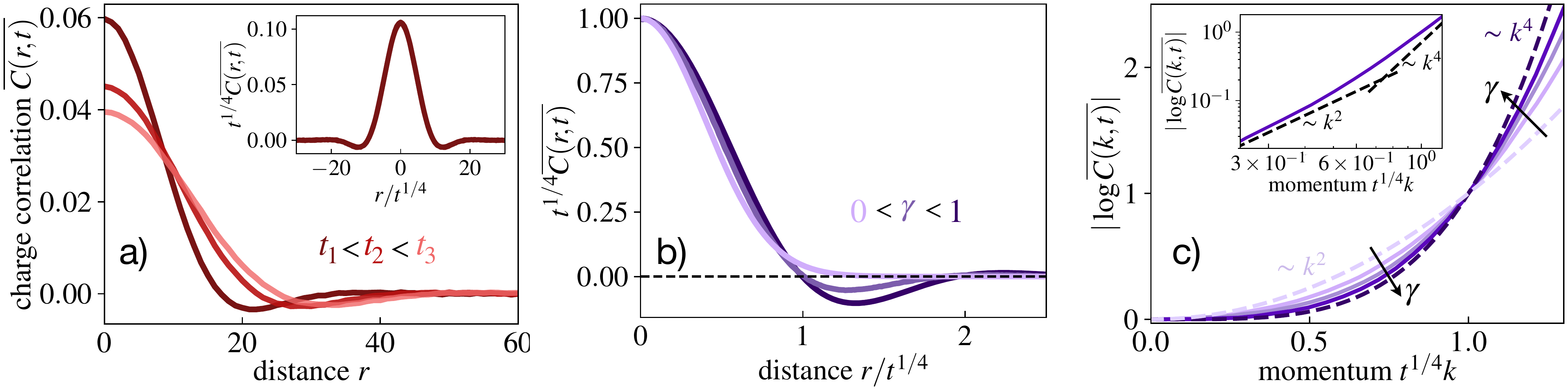}
\caption{\textbf{Hydrodynamic phase mixing.} \textbf{a)} The profile of the spin correlations at different times for a probability $\gamma \neq 0$ of random circuit updates that conserve the  dipole-moment of the spin pattern. \textit{Inset}: A scaling collapse of the profiles determines the dynamical exponent $z=4$. \textbf{b)} The scaling function associated to some $0\leq \gamma \leq 1$ is neither the Gaussian, that is associated to hard core tracer diffusion, nor the fundamental solution of the dipole-conserving hydrodynamic equation \eq{eq:4.0.1}. It is instead given by a convolution of the two. Here, the scaling functions are normalized such that their value at $r=0$ is  equal to one. \textbf{c)} Logarithm of the Fourier transform $C(k,t)$ of the real space correlations $C(r,t)$. The momentum axis has been rescaled independently for different $\gamma$, defined in such a way that all curves coincide at $t^{1/4}k=1$, $|\log \overline{C(k,t)}|=1$. This allows us to directly compare the relative strengths of the $(kt^{1/4})^2$-- and $(kt^{1/4})^4$-- contributions to $|\log \overline{C(k,t)}|$ for different $\gamma$. We see that increasing the probability $\gamma$ of dipole-conserving spin updates leads to an increasing weight of the $(t^{1/4}k)^4$-term in \eq{eq:4.2.3}. Inset: On a double logarithmic scale the crossover from $(kt^{1/4})^2$-- to $(kt^{1/4})^4$-- behavior becomes visible.}
\label{fig:10}
\end{figure*}

\subsection{$m=0$: Diffusion}
For $m=0$, under rescaling space and time in \eq{eq:4.0.3} according to $k \rightarrow k/\lambda$, $t \rightarrow \lambda^z t = \lambda^2 t$, we obtain
\begin{equation} \label{eq:4.1.1}
\begin{split}
C(k,t) \sim &\exp\Bigl\{-D_sk^2t - \sqrt{Dt}\, k^2/\lambda\Bigr\} \\ 
&\xrightarrow{\lambda \rightarrow \infty} \exp\Bigl\{-D_sk^2t \Bigr\}.
\end{split}
\end{equation}
This implies that the dynamical exponent is $z=2$.
Therefore, if only the total spin of the pattern is conserved the correlations $C(k,t)$ are described by conventional diffusion as expected. We verify this result numerically in $tJ$ -- like random unitary circuits with charge-conserving two-spin gates, see \figc{fig:8}{a}. The mean squared displacement $\sigma^2(t)$ of the resulting real space correlations $\overline{C(r,t)}$ indeed scales diffusively $\sigma^2(t)\sim t$ at long times as shown in \figc{fig:8}{b}.

\subsection{$m\geq 2$: Tracer diffusion}
On the other hand, rescaling $k \rightarrow k/\lambda$, $t \rightarrow \lambda^z t = \lambda^4 t$ for $m\geq 2$ in \eq{eq:4.0.3} yields
\begin{equation}
\begin{split}
C(k,t) \sim &\exp\Bigl\{-D_sk^{2m+2}t/\lambda^{2m-2} - \sqrt{Dt}\, k^2 \Bigr\} \\ 
&\xrightarrow{\lambda \rightarrow \infty} \exp\Bigl\{- \sqrt{Dt}\, k^2 \Bigr\},
\end{split}
\end{equation}
and the dynamical exponent is $z=4$.
Thus, if all moments of the spin pattern up to at least the quadrupole moment are conserved the long-time correlations remain dominated by the anomalously slow tracer motion of \eq{eq:3.1.1}. Again, we verify this numerically by computing the mean-squared displacement $\sigma^2(t) = \sum_r r^2 \, \overline{C(r,t)}$ of a random time evolution with $m-$pole conserving spin interactions, see \figc{fig:8}{b}. In practice, we have to make sure to avoid localization of the spin pattern dynamics due to a strong fragmentation of the Hilbert space into disconnected subsectors, which can occur for all $m\geq 1$~\cite{Sala19,khemani20192d,Rakovszky20}. This is achieved by choosing spin-gates of sufficient range, which ensures ergodicity of the spin dynamics. As expected, the system is described by $z=4$ subdiffusive tracer dynamics in the case of quadrupole conservation $m=2$, i.e. $\sigma^2(t) \sim \sqrt{t}$, see \figc{fig:8}{b}.

\subsection{$m=1$: Hydrodynamic phase coexistence}
For the special case of $m=1$, both terms in the exponent of $C(k,t)$ are equally relevant under the rescaling $k \rightarrow k/\lambda$, $t \rightarrow \lambda^4 t$,
\begin{equation}
C(k,t) \xrightarrow{\lambda \rightarrow \infty} \exp\Bigl\{-D_sk^{4}t - \sqrt{Dt}\, k^2 \Bigr\}.
\end{equation}
The correlation function $C(k,t)$ is thus subject to a competition between two inequivalent dynamical processes that both have $z=4$ but that have different forms of their respective scaling function. Notably, although $C(k,t)=C(kt^{1/4})$ is a function of $kt^{1/4}$, it can not be written in terms of some universal scaling function $\mathcal{K}(\cdot)$ that is independent of microscopic details. Instead,
\begin{equation}
C(k,t) = \mathcal{K}\Bigl(k(Dt)^{1/4},D_s/\sqrt{D}\Bigr),
\end{equation}
i.e. the form of $C(kt^{1/4})$ depends non-trivially on the ratio $D_s/\sqrt{D}$ which determines the mixture of the two universal processes. Specifically, we can express
\begin{equation} \label{eq:4.2.3}
\log C(k,t) = - (D_s+\sqrt{D}) \Bigl[ \mu\, (kt^{1/4})^4 + (1-\mu)\, (kt^{1/4})^2\Bigr], 
\end{equation}
where $\mu=\mu(D_s/\sqrt{D})=\frac{D_s}{\sqrt{D}}(1+\frac{D_s}{\sqrt{D}})^{-1}$ and $0\leq \mu \leq 1$. The specific mixture, and thus the long time and length scale profile of the spin correlations, is sensitive to microscopic details of the time evolution, reminiscent of UV-IR-mixing~\cite{you2020_uvir,seiberg2020_uvir,gorantla2021_uvir,you2021_uvir,hart2021_experimental,sala2021_dynamics,hart2022_hidden}. This results in a continuously varying hydrodynamic universality class controlled by the microscopic mixing parameter $\mu$. Here, we identify a hydrodynamic universality class with both the dynamical exponent and the scaling function.

We confirm these theoretical considerations numerically in \fig{fig:10}, where we consider a random unitary time evolution with dipole-conserving dynamics within the spin pattern, see \figc{fig:8}{a}. To ensure ergodicity, we use dipole-conserving spin updates ranging over eight sites. For any given probability $\gamma$ at which non-trivial spin pattern rearrangements occur the dynamical exponent is $z=4$, as demonstrated by the scaling collapse of $\overline{C(r,t)}$ evaluated at different times in \figc{fig:10}{a}. However, varying this probability $\gamma$ effectively controls the ratio $D_s/\sqrt{D}$ and leads to different scaling functions as shown in \figc{fig:10}{b}. In particular, the limiting distributions are a Gaussian for $\gamma=0$ and the dipole-conserving hydrodynamic scaling function of \eq{eq:4.0.2} (for $m=1$) as $\gamma \rightarrow 1$. In addition, we numerically compute the Fourier transform $C(k,t)$ of the correlation profile to verify the prediction of \eq{eq:4.2.3}. \figc{fig:10}{c} shows that increasing the rate $\gamma$ of the spin dynamics leads to an increasing contribution of the $(kt^{1/4})^4-$term to $\log C(k,t)$.

\begin{figure}[t]
\centering
\includegraphics[trim={0cm 0cm 0cm 0cm},clip,width=0.95\linewidth]{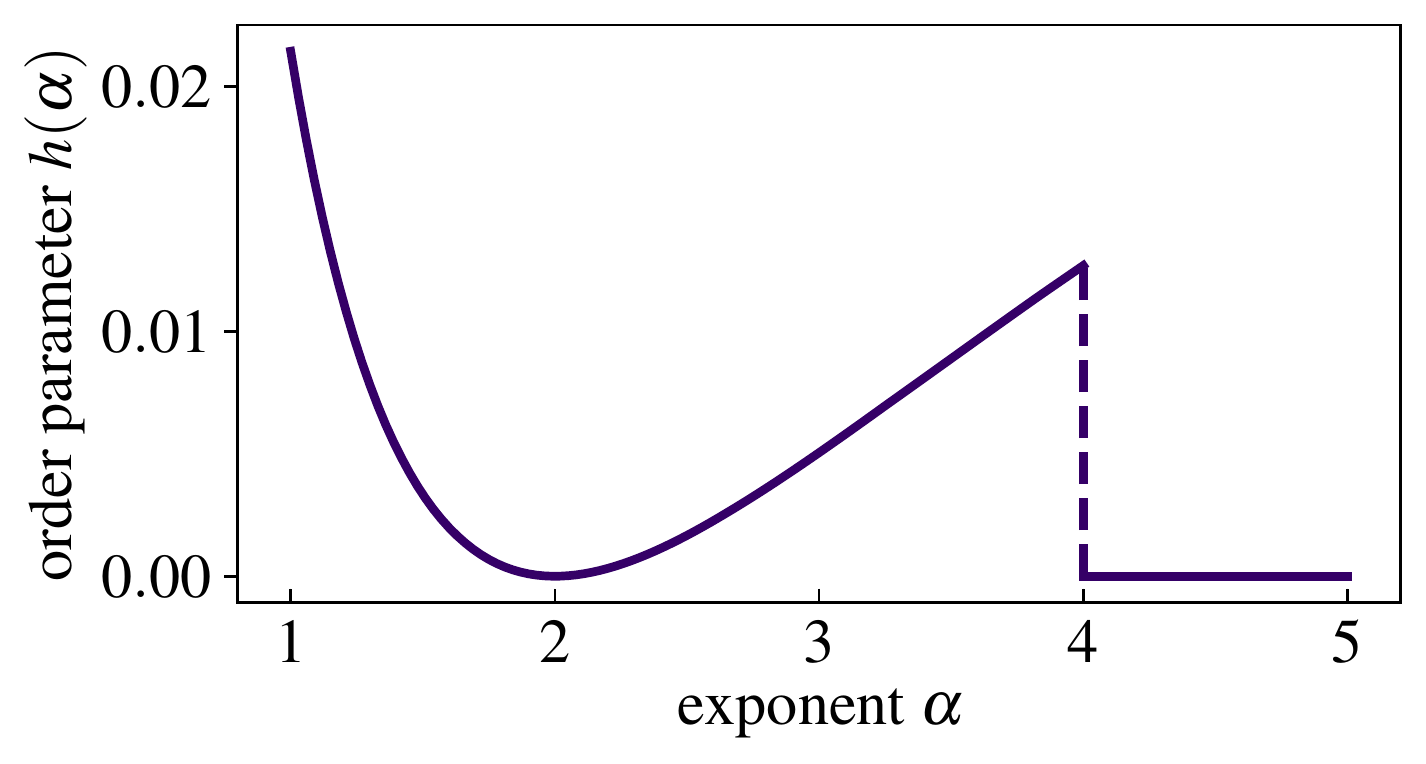}
\caption{\textbf{Hydrodynamic order parameter.} The order parameter $h(\alpha)$ of \eq{eq:4.2.7} quantifies the deviation of the charge correlation profile from a Gaussian. $\alpha$ labels the exponent of the spin pattern's internal dynamics, see \eq{eq:4.2.4}. There is a clear discontinuity at $\alpha=4$, where long-time dynamical correlations switch between Gaussian tracer motion and a non-Gaussian dipole-conserving profile. As a consequence, $\alpha = 4$ can be interpreted as a first order dynamical transition, which allows for phase coexistence of distinct hydrodynamic universality classes.}
\label{fig:11}
\end{figure}

The arbitrary mixing of two distinct dynamical scaling functions as in \eq{eq:4.2.3} can be viewed as phase coexistence of two hydrodynamic phases, in analogy with more conventional phase coexistence occuring at first order equilibrium transitions. To make this analogy more tangible, let us imagine a situation in which the dynamics of the spin pattern is given by
\begin{equation} \label{eq:4.2.4}
F_\alpha(k,t) = \frac{1}{\sqrt{2\pi}} \, \exp(-D_s \, |k|^{\alpha} t),
\end{equation}
now with some general exponent $0<\alpha<\infty$ whose value is governed by some underlying model (e.g. through the power-law decay of a long-ranged spin term in the constrained $tJ$ -- like models). Using that $C_\alpha(k,t)=F_\alpha(k,t)G_{tr}(k,t)$, we obtain the dynamical exponent $z$ as a function of $\alpha$:
\begin{equation} \label{eq:4.2.5}
z(\alpha) = 
\begin{cases}
\alpha, \text{ for } \alpha<4 \\
4, \text{ for } \alpha \geq 4
\end{cases}.
\end{equation}
In particular, $\alpha>4$ corresponds to Gaussian tracer motion while $\alpha <4$ is associated with non-Gaussian scaling functions. In addition, for $\alpha \neq 4$ we can always write the real space correlations $C_\alpha(r,t)$ as 
\begin{equation} \label{eq:4.2.6}
C_\alpha(r,t) = (Dt)^{-1/4} \mathcal{F}_\alpha \bigl(r(Dt)^{-1/4}\bigr),
\end{equation}
with a normalized universal scaling function $\int dx\, \mathcal{F}_\alpha(x) = 1$.
If we thus consider $\alpha=4$ to separate a Gaussian and a non-Gaussian dynamical phase, we can accordingly define an order parameter that quantifies the non-Gaussianity of the scaling function for a given $\alpha$ via
\begin{equation} \label{eq:4.2.7}
\begin{split}
&h(\alpha) := \min_{\lambda > 0} \int dx\, \Bigl( \lambda\,\mathcal{F}_\alpha(\lambda x) - \frac{1}{\sqrt{\pi}}\exp(-x^2) \Bigr)^2 = \\
&=
\begin{cases}\frac{1}{2\pi}\min\limits_{\lambda>0} \int dk\, \Bigl( e^{-(|k|/\lambda)^\alpha}-e^{-k^2/4} \Bigr)^2, \quad \alpha < 4 \\
0, \quad \alpha > 4
\end{cases}.
\end{split}
\end{equation}
We have evaluated $h(\alpha)$ numerically in \fig{fig:11}, where we see a clear discontinuity at $\alpha=4$. The central property $\Delta h(\alpha=4) > 0$ can also be demonstrated analytically. In particular, the variance of the $\alpha \rightarrow 4^-$ scaling function vanishes, as opposed to a Gaussian~\cite{feldmeier2020anomalous}. This jump in the order parameter suggests that we can indeed interpret the point $\alpha=4$ as a first order dynamical transition, with \eq{eq:4.2.3} describing the Gaussian/non-Gaussian phase mixture.

\section{Conclusion \& Outlook}
In this work, we investigated the emergent hydrodynamics of $tJ$ -- like many-body systems in one dimension with constrained spin interactions. We found that for chaotic, thermalizing systems the dynamical spin correlation function at infinite temperature is given by a convolution of the dynamics of the underlying spin pattern and the tracer motion of hard core particles. In $tJ_z$ -- like systems all multipole moments of the spin pattern are constants of motion and spin correlations are given by tracer dynamics alone. This allowed us to demonstrate the emergence of subdiffusion with dynamical exponent $z=4$ in several random circuit lattice models that feature a (effective) constant spin pattern. Using results from the theory of tracer motion we provided expressions for the full long-time profile of dynamical spin correlations in these models. It will be interesting to see in the future whether additional one-dimensional systems fall under this dynamical universality class.
We also remark that although we mostly focused on situations with a conserved $\sigma_i = \pm 1$ spin pattern, our results generalize to patterns of higher effective spin $\sigma_i = -|S|,...,|S|$. In such an instance, all odd power spin density correlations of the form
\begin{equation} \label{eq:6.1}
\Braket{ \bigl(\hat{S}^z_x(t)\bigr)^{2n+1} \bigl(\hat{S}^z_0(0)\bigr)^{2n+1} },
\end{equation}
with integer $n$, reduce to the same tracer process up to a global prefactor. We emphasize that our results apply to infinite temperature correlations. Whether extensions to finite temperatures are possible depends sensitively on whether the average charge of the pattern vanishes also at finite temperatures for a given model.

We further established a connection to integrable $tJ_z$ -- like quantum systems which feature conserved spin patterns and a time evolution that is independent of the pattern. By mapping to a tracer problem of ballistically moving particles, we were able to reproduce the spin diffusion constant of the folded XXZ chain and provided its full long-time correlation profile. The characteristic staggered oscillations of the resulting correlation profile can be verified in quantum simulation experiments on XXZ chains already at moderate anisotropy. We further point out a connection to anomalous current correlations recently reported for the XXZ chain, the XNOR circuit, and for a deterministic classical automaton, Refs.~\cite{gopalakrishnan2022_anomalous,krajnik2022_anomalous}. Ref.~\cite{gopalakrishnan2022_anomalous} provides a picture in which such correlations are effectively due to the tracer motion of a single domain wall. We expect that the same arguments, and thus the presence of anomalous correlations, apply in any system with a conserved pattern, which includes Ref.~\cite{krajnik2022_anomalous}.

Moreover, it is interesting to study non-integrable Hamiltonian quantum systems with conserved patterns, e.g. via the $tJ_z$ -- model at finite $J_z$ or via adding diagonal interactions to the folded XXZ model. For example, a stochastic spin chain closely related to the folded XXZ model is the so-called Ising-Kawasaki model~\cite{grynberg,vinet}, taken in a particular limit: the Hamiltonian $H_{\rm fXXZ}$ is supplemented by nearest-neighbor (NN) and next-nearest-neighbor (NNN) Ising terms. The name derives from the fact that the quantum Hamiltonian is obtained from the Markov operator of a classical Ising chain undergoing Kawasaki magnetization-conserving dynamics. The NNN Ising coupling does not commute with $H_{\mathrm{fXXZ}}$, leading to integrability breaking in many sectors~\cite{yang2020_strict} while preserving the pattern conservation described above.
Such systems should fall under the same universality class as the generic random circuits considered in this work and we expect them to exhibit $z=4$ subdiffusive hard core tracer dynamics. A version of the folded XXZ model in which integrability is broken by noise was recently studied in Ref.~\cite{deNardis2021_subdiff}, concluding $z=4$ as well.

In addition, we investigated the dynamics of generic systems in which only a finite number of multipole moments of the spin pattern remains conserved. We found that the characteristics of tracer dynamics survive if at least the multipole moments up to and including the quadrupole moment are constant. Intriguingly, if only the moments up to the dipole are conserved there emerges a special scenario in which spin correlations are subject to a competition between two hydrodynamic processes with dynamic exponent $z=4$ but different scaling functions. The resulting shape of the long-time correlations are then susceptible to microscopic details of the time evolution and we find the situation to be reminiscent of the coexistence of different hydrodynamic phases at a first order transition. How such a scenario may qualitatively arise in systems other than the ones considered here is an interesting open question for future research.\\

\textbf{\textit{Acknowledgments.}}--
We thank Alvise Bastianello, Sarang Gopalakrishnan, Oliver Hart, Gabriel Longpr\'e, Frank Pollmann, Tibor Rakovszky, Pablo Sala, Stéphane Vinet and Philip Zechmann for insightful discussions.
We acknowledge support from the Deutsche Forschungsgemeinschaft (DFG, German Research Foundation) under Germany’s Excellence Strategy-EXC-2111-390814868, TRR80 and DFG grant No. KN1254/2-1, No. KN1254/1-2, the European Research Council (ERC) under the European Union’s Horizon 2020 research and innovation programme (grant agreement No. 851161), as well as the Munich Quantum Valley, which is
supported by the Bavarian state government with funds from the Hightech Agenda
Bayern Plus. We thank the Nanosystems Initiative Munich (NIM) funded by the German Excellence Initiative and the Leibniz Supercomputing Centre for access to their computational resources. 
W.W.-K. was funded by a Discovery Grant from NSERC, a Canada Research Chair, and a grant from the Fondation Courtois.
Matrix product state simulations were performed using the TeNPy package~\cite{hauschild2018_tenpy}.

\textbf{\textit{Data and materials availability.}}-- Data analysis and simulation codes are available on Zenodo upon reasonable request~\cite{zenodo}.

\appendix 
\section{Strength of staggered oscillations} \label{sec:App1}
In this appendix we derive the exact analytical expression \eq{eq:3.3.15} for the constant $C_0$ defined in \eq{eq:3.3.10}. This constant determines the strength of the staggered oscillations on top of the Gaussian enveloping shape for the dynamical spin correlation profile in both the random XNOR circuit and the folded XXZ model, \textit{cf.} \eqs{eq:3.3.19}{eq:3.4.10}.
We recall that according to \eq{eq:3.3.12}, $C_0$ can be written in the domain wall picture as
\begin{equation} \label{eq:A.1.1}
C_0 = \Braket{\Bigl( \hat{\tilde{\kappa}}_0^{(single)} \Bigr)^2} \sum_r \Braket{ \hat{\tilde{\kappa}}_{2r}^{(single)} }^\prime,
\end{equation}
where the expectation value $\braket{\cdot}^\prime$ is taken with respect to an ensemble where a domain wall with positive charge $\tilde{\kappa}_0^{(single)}=1$ is fixed to sit at bond $0$. We have already evaluated $\Braket{\Bigl( \hat{\tilde{\kappa}}_0^{(single)} \Bigr)^2} = 1/6$ in the main text and so we focus on the correlation function 
\begin{equation} \label{eq:A.1.2}
\Braket{\hat{\mathcal Q}_A}^\prime := \sum_r \Braket{ \hat{\tilde{\kappa}}_{2r}^{(single)} }^\prime.
\end{equation}
$\Braket{\hat{\mathcal Q}_A}^\prime$ counts the $A$-sublattice charge of single domain walls provided a positive domain wall is located at the origin. The sublattice charge \eq{eq:A.1.2} refers only to \textit{single} domain walls and is conserved in the time evolution. In particular, $\Braket{\hat{\mathcal Q}_A}^\prime$ can be calculated entirely within the ensemble of all possible conserved domain wall patterns, see \fig{fig:5}, where mobile domain wall pairs have already been removed. This is our approach in the following.

We first recall that the ensemble of possible conserved patterns in the domain wall picture is subject to an exclusion principle of nearest-neighbor domain walls, see \fig{fig:5}. That is, a bond with a domain wall must have two aligned bonds as its neighbors. For simplicity and without loss of generality we can always assume the leftmost domain wall in the system to have positive charge, fixing a $\mathds{Z}_2$ degree of freedom for the staggering. The exclusion property then gives rise to a Fibonacci sequence for the number $N_p(\ell)$ of possible conserved pattern configurations for a number $\ell$ of bonds:
\begin{equation} \label{eq:A.1.3}
N_p(\ell) = N_p(\ell-1) + N_p(\ell-2).
\end{equation}

Given that a (positive) domain wall is fixed to sit at the origin, let us move to the right of the origin and derive the probability $p(r)$ to encounter the \textit{next} domain wall exactly at a distance $r$. We find $p(1)=0$ due to the exclusion principle and 
\begin{equation}\label{eq:A.1.4}
p(r\geq 2) = \frac{N_p(\ell-r)}{N_p(\ell)} \xrightarrow{\ell \rightarrow \infty}\varphi^{-r},
\end{equation}
with the golden ratio $\varphi=\frac{1+\sqrt{5}}{2}$. As required, $\sum_{r=1}^{\infty}p(r) = 1/(\varphi^2-\varphi)=1$. Let us then further derive the probability $p(A|A)$ that as we go to right from a domain wall located on the $A$ sublattice, the next domain wall we encounter is again located on the $A$ sublattice,
\begin{equation} \label{eq:A.1.5}
p(A|A) = \sum_{n=1}^{\infty} p(2n) = \varphi^{-2}\sum_{n=0}^{\infty}\varphi^{-2n} = \frac{1}{\varphi^2-1}=\frac{1}{\varphi},
\end{equation}
where we used the defining equation of the golden ratio in the last equality.
From this result, we also obtain the probability $p(B|A)=1-p(A|A)$ to find the next domain wall we encounter on the $B$ sublattice, given that the previous one is located on the $A$ sublattice. Similarly, we have $p(B|B)=p(A|A)$ and $p(A|B)=p(B|A)$.

We now take into account that the charges of the domain walls have to be perfectly anticorrelated, i.e., if there is a positive domain wall at the origin, the next domain wall we encounter to the right must have negative charge. Therefore, moving to the right from the positive domain wall at the origin, we can determine the charge of the next domain wall that we find at an $A$ sublattice bond by counting the number of $B$ sublattice domain walls in between. The probability $p(A,n_B=0,A)$ to find no other domain wall between two consecutive $A$ domain walls is $p(A,n_B=0,A)=p(A|A)$. The probability to find a number $n_B\geq 1$ of $B$ sublattice domain walls between two consecutive $A$ domain walls is given by
\begin{equation} \label{eq:A.1.6}
\begin{split}
p(A,n_B\geq 1,A) &= p(A|B) \bigl[p(B|B))\bigr]^{n_B-1} p(B|A) = \\
&= \bigl[1-p(A|A)\bigr]^2 \bigl[p(A|A)\bigr]^{n_B-1}.
\end{split}
\end{equation}
If the number of $B$ sublattice domain walls between two consecutive $A$ domain walls is even, the two $A$ domain walls have opposite charge. If the number of $B$ sublattice domain walls between two consecutive $A$ domain walls is odd, they have equal charge. Therefore, the probability to find two consecutive $A$ sublattice domain walls with opposite charge is given by
\begin{equation} \label{eq:A.1.7}
\begin{split}
P_- &= \sum_{n=0}^{\infty} p(A,n_B=2n,A)  \\
&= p(A|A) + \bigl[1-p(A|A)\bigr]^2 p(A|A) \sum_{n=0}^{\infty} \bigl[p(A|A)\bigr]^{2n} \\
&= p(A|A) + \bigl[1-p(A|A)\bigr]^2 \frac{p(A|A)}{1-\bigl[p(A|A)\bigr]^2} = \frac{2}{\varphi^2},
\end{split}
\end{equation}
where in the last equality we inserted \eq{eq:A.1.5}. Accordingly, the probability to find two consecutive $A$ sublattice domain walls with equal charge is given by
\begin{equation} \label{eq:A.1.8}
P_+ = 1-P_- = 1-\frac{2}{\varphi^2}.
\end{equation}
We note that $P_- \approx 0.764 > 0.5$, i.e.\ two consecutive $A$ domain wall charges are anticorrelated. This is a `remainder' of the perfect anticorrelation between two consecutive domain walls irrespective of the sublattice.

Let us finally take a randomly chosen conserved domain wall pattern configuration with a positive domain wall fixed at the origin and consider the $r$-th $A$ sublattice domain wall to the right from the origin. The charge of the $r$-th $A$ domain wall is then determined by a sequence of exactly $r$ pairs of two consecutive $A$ domain walls. The probability to find $r^\prime \leq r$ anticorrelated consecutive $A$ domain wall pairs in this sequence is $\binom{r}{r^\prime}(P_-)^{r^\prime}(P_+)^{r-r^\prime}$, where the binomial coefficient accounts for the reordering of the anticorrelated pairs in the sequence. Since the domain wall at the origin is positive, a sequence containing $r^\prime$ anticorrelated consecutive domain wall pairs implies a charge of $(-1)^{r^\prime}$ for the $r$-th $A$ domain wall. Perfoming a sum over the possible number $0\leq r^\prime \leq r$ of anticorrelated pairs in the sequence and summing over the charge contributions from all $A$ sublattice domain walls we obtain
\begin{equation} \label{eq:A.1.9}
\begin{split}
\Braket{\hat{\mathcal Q}_A}^\prime
&= 1 + 2 \sum_{r=1}^{\infty}\sum_{r^\prime=0}^{r} \binom{r}{r^\prime} (-1)^{r^\prime}(P_-)^{r^\prime}(P_+)^{r-r^\prime} = \\
&= 1 + 2 \sum_{r=1}^{\infty} (P_+-P_-)^r = \frac{P_+}{P_-} = \frac{\varphi^2}{2}-1.
\end{split}
\end{equation}
In the first line of the above equation, the first term of unity is due to the positive contribution of the positive charge fixed at the origin, while the factor of two is due to symmetric contributions from right and left of the origin. According to \eq{eq:A.1.2}, \eq{eq:A.1.9} yields the correlation function $\sum_r \Braket{ \hat{\tilde{\kappa}}_{2r}^{(single)} }^\prime$. Inserting into \eq{eq:A.1.1} finally we obtain
\begin{equation}
C_0 = \frac{1}{6}\Bigl( \frac{\varphi^2}{2}-1 \Bigr) = \frac{1}{12}\bigl(\varphi-1\bigr),
\end{equation}
completing our proof.

\bibliography{tJreferences}

\end{document}